# Fluid-solid transitions in photonic crystals of soft, thermoresponsive microgels

M. Hildebrandt,[1] D. Pham Thuy,[1] J. Kippenberger,[1] T. L. Wigger,[1] J. E. Houston,[2] A. Scotti,[3] and M. Karg[1]*

[1]Institut für Physikalische Chemie I: Kolloide und Nanooptik, Heinrich-Heine-Universität Düsseldorf, Universitätsstraße 1, D-40225 Düsseldorf, Germany

E-Mail: karg@hhu.de

[2]European Spallation Source ERIC, Box 176, SE-221 00 Lund, Sweden

[3]Institute of Physical Chemistry, RWTH Aachen University, Landoltweg 2, 52056 Aachen, Germany

**Abstract**

Microgels are often discussed as well-suited model system for soft colloids. In contrast to rigid spheres, the microgel volume and, coupled to this, the volume fraction in dispersion can be manipulated by external stimuli. This behavior is particularly interesting at high packings where phase transitions can be induced by external triggers such as temperature in case of thermoresponsive microgels. A challenge, however, is the determination of the real volume occupied by these deformable, soft objects and consequently, to determine the boundaries of the phase transitions. Here we propose core-shell microgels with a rigid silica core and a crosslinked, thermoresponsive poly-*N*-isopropylacrylamide (PNIPAM) shell with a carefully chosen shell-to-core size ratio as ideal model colloids to study fluid-solid transitions that are inducible by millikelvin changes in temperature. Specifically, we identify the temperature ranges where crystallization and melting occur using absorbance spectroscopy in a range of concentrations. Slow annealing from the fluid to the crystalline state leads to photonic crystals with Bragg peaks in the visible wavelength range and very narrow linewidths. Small-angle X-ray scattering is then used to confirm the structure of the fluid phase as well as the long-range order, crystal structure and microgel volume fraction in the solid phase. Thanks to the scattering contrasts and volume ratio of the cores with respect to the shells, the scattering data do allow for form factor analysis revealing osmotic deswelling at volume fractions approaching and also exceeding the hard sphere packing limit.

## 1. Introduction

Microgels and nanogels are composed of chemically and/or physically crosslinked polymer chains and are swollen by large amounts of solvent under good and typically even bad solvent conditions.[1-2] With this large solvent content and their soft polymeric network structure the physical properties of microgels are complex featuring characteristics that are common for colloids, surfactants and macromolecules.[3-8] Depending on the polymer composition, microgels can react to different external stimuli like temperature, pH or ionic strength.[9-11] The most prominent example of temperature responsive microgels is based on poly-*N*-isopropylacrylamide (PNIPAM). PNIPAM microgels undergo a pronounced volume phase transition (VPT) with an extend and transition temperature of 32°C or higher depending on the crosslinker density.[12-14] This behavior is related to the lower critical solution temperature (LCST) of PNIPAM in water.[15-16] The precipitation of PNIPAM at temperatures above the LCST is also the basis for the rather straightforward synthesis of related microgels by (seeded) precipitation polymerization.[17-18] The resulting microgels typically possess an inhomogeneous internal network structure resembling a gradient in crosslinking and consequently a polymer density that is decreasing in the outer periphery of the microgels. This gradient is the result of the faster consumption of the crosslinker during the precipitation polymerization and leads to a higher crosslinked core and a loose outer shell that also contains dangling chains.[19-21] This structure was also found for the polymeric shell of core-shell microgels that contain rigid nanoparticle cores such as gold[22] and silica.[23] In recent years the structural evolution of microgels in dense packings, either in 2-dimensional confinement or 3-dimensional samples in bulk dispersion, has attracted significant interest.[1, 24-25] Despite their softness, microgels can (self-) assemble into highly ordered structures and, for example, crystalline phases similar to hard spheres were observed.[5, 12-13, 26-31] In contrast to rigid colloids, responsive microgels allow for external control of the microgel volume fraction due to their VPT behavior, *i.e.* for a given number concentration, $N$, for example temperature can be used to alter the volume fraction and thereby induce phase transitions.[28, 32] This behavior renders microgels particularly interesting for fundamental studies on crystallization, crystal melting and jamming. Surprisingly, little is known about the minimal temperature change needed to induce the fluid/solid and solid/fluid transitions.[28, 30] A challenge related to structural investigations at high volume fractions is the deswelling and faceting that alters the form factor of the microgels in dependence of their concentration.[33-40] This behavior in combination with the inhomogeneous internal network structure and the high solvent contents hamper the determination of the volume fraction. The easier accessible quantity is the generalized volume fraction that considers the microgel volume in the dilute, non-interacting state.[32, 41] In-situ studies on phase transitions consequently require knowledge of the form as well as structure factor in dense packings that typically requires deuteration strategies and complex contrast variation using X-ray and neutron scattering.[20, 34]

In this work, we use silica-PNIPAM core-shell (CS) microgels to study fluid-solid transitions in dense packings. Our aims are (i) to determine transition temperatures where fluid-solid transitions occur, (ii) to study the structure and long-range order of both phases and (iii) to determine the microgel size and volume fraction in the concentration range where osmotic deswelling is expected to alter the microgel form factor. To do so, we use a combination of absorbance spectroscopy and synchrotron SAXS measurements for samples of various concentrations in the regime of dense packing. Importantly, the shell-to-core size ratio of the CS microgels was chosen to provide contrast for both, the shell and the core in SAXS investigations. This allows for the direct determination of structure factors as well as form factors at high volume fractions from single, individual scattering profiles. In the solid regime, we use Bragg peak analysis to study the crystal structure and domain size.

## 2. Experimental Section

**Chemicals**

L-arginine (PanReacAppliChem, ≥ 99%), cyclohexane (Fisher Scientific, analytical reagent grade), tetraethyl orthosilicate (TEOS, Sigma Aldrich, 98%), 3-(trimethoxysilyl)propyl methacrylate (MPS, Sigma Aldrich, 98%), *N,N*-methylenebis(acrylamide) (BIS, Sigma-Aldrich, 99%), potassium persulfate (PPS, Sigma-Aldrich, 99,0%) and heavy water ($D_2O$, Sigma Aldrich, 99.9%) were used as received without further purification.

*N*-isopropylacrylamide (NIPAM, TCI, >98%) was recrystallized from cyclohexane prior to use. Water was always used in ultra-high purity provided by a Milli-Q system (Merck Millipore) with a resistivity of 18.2 MΩ cm.

## Synthesis of silica cores

Silicon dioxide seed particles were synthesized by an interphase-mediated condensation reaction,[42] where 82 mg (0.47 mmol) of L-arginine were dissolved in 78 mL of water in an 100 ml Erlenmeyer flask and an organic phase was added by the addition of 4.05 mL of cyclohexane. The reaction mixture was equilibrated for 30 minutes at 60°C in an oil bath and stirred only gently in order to prevent the aqueous and organic phase from mixing. Afterwards, 4.95 mL (22.2 mmol) of TEOS were added to the organic phase and the reaction was allowed to proceed for 72 h. The resulting silica particles were then surface-functionalized with MPS to increase the hydrophobicity as needed for the later precipitation polymerization. To do so, the aqueous phase was used without any purification and 9 mL of cyclohexane were added to the dispersion. After equilibration for 30 minutes at 60°C, 90 µL of MPS were added to the organic phase. After stirring for 24 h, the aqueous phase was separated and the mass content of the dispersion was determined. The obtained MPS-functionalized seeds were used without further purification.

## Synthesis of CS microgels

CS microgels were synthesized by free-radical precipitation polymerization using 4000 mg (35.4 mmol) NIPAM monomer, 817 mg (5.3 mmol) of BIS as crosslinking comonomer and 122 mg MPS-functionalized seeds dispersed in 600 mL of water. The reaction medium was degassed with nitrogen for 45 minutes at a temperature of 70°C. Then, 50 mg (0.19 mmol) of PPS dissolved in 1 mL of water were added in one shot to initiate the polymerization. The reaction was conducted for 6 h. The CS microgels were purified and concentrated by three centrifugation and redispersion (water) steps at 10000 rcf for 3 h followed by dialysis against pure water for at least seven days with multiple water exchanges.

## Sample preparation

CS microgel dispersions with defined mass contents used for absorbance spectroscopy and SAXS measurements were prepared by freeze-drying the respective amount of a stock dispersion with known mass content and redispersion of the microgels in the respective volume of $D_2O$ or $H_2O$. CS microgel dispersions with only low mass contents were prepared by dilution of a stock dispersion. Dense samples at high concentrations were prepared in 0.2 mm x 4.00 mm x 10 mm rectangular, flat-wall capillaries (VitroTubes). The respective CS microgel dispersions were sucked into the capillaries by applying a small reduced pressure from one side of the capillary. Afterwards, the capillaries were flame-sealed with a hydrogen torch. After sample preparation, the samples were annealed using a temperature cycle. First, the samples were heated from 20 °C to 50 °C with a rate of 1.5 °C/h. Then the temperature was kept constant at 50 °C for 1 h. Finally, the samples were cooled back to 20 °C with a rate of 1.5 °C/h. For SAXS measurements above 20 °C, we prepared CS microgel dispersions in 1 mm round capillaries (WJM Glas) and sealed with epoxy glue. A sample list is given in **Table S1** in the supporting information.

**SAXS**

Synchrotron SAXS experiments were performed on the CoSAXS beamline at the MAX IV synchrotron in Lund (Sweden). The instrument was equipped with an Eiger2 4M detector exhibiting a sensitive area of 155.1 × 162.2 mm$^2$ with total pixel sizes of 75 × 75 µm$^2$. SAXS measurements were recorded with acquisition times of 10 ms. The energy of the X-ray beam was 12.4 keV and the sample-to-detector distance was set to 6.85 m resulting in an effective $q$-range from 0.015 nm$^{-1}$ to 0.5 nm$^{-1}$ for the measurements performed at 20 °C. For the experiments performed at 40 °C the sample-to-detector distance was set to 11.04 m resulting in an effective $q$-range from 0.01 nm$^{-1}$ to 0.5 nm$^{-1}$. Detector images were radially averaged and the resulting SAXS profiles were background corrected for D$_2$O.

In addition, SAXS measurements on selected samples were performed on a Xeuss 2.0 (XENOCS) equipped with a Pilatus3R 300K detector exhibiting a sensitive area of 83.8 × 106.5 mm$^2$ with total pixel sizes of 172 x 172 µm$^2$. The sample-to-detector distance was set to 1.2 m and the energy of the X-ray beam was 8.048 keV leading to an effective $q$-range of 0.03 nm$^{-1}$ to 2 nm$^{-1}$. All measurements were performed in order to extract scattering intensities in absolute units with 1 mm glassy carbon as reference. Scattering profiles were recorded with acquisition times of 3600 s. The resulting SAXS profiles were corrected for the respective solvent as background. CS microgel dispersions with mass contents of 10 wt% were    and SiO$_2$ particles prior to encapsulation were prepared in 1 mm round capillaries (WJM Glas).

Radial averaged scattering profiles were analyzed with the SasView software .[43]The 2D SAXS patterns were simulated with the software Scatter by Förster and Apostol.[44]

**Absorbance spectroscopy**

UV-Vis-NIR absorbance spectroscopy was performed on a SPECORD S600 (Analytik Jena) equipped with a temperature-controlled sample changer. We want to note that we measured the exact same samples by absorbance spectroscopy as by synchrotron SAXS.

**Angle-dependent reflectance spectroscopy**

Reflectance spectroscopy was performed with a home-built setup consisting of a tungsten-lamp as light source and the Flame VIS-NIR Spectrometer (Ocean Insight) as detector. Both, detector and the light source were equipped with visible-NIR fibers to ensure free movement of the detector and light source in a goniometer-like fashion. The sample holder and fiber guides were 3D printed using a Prusa MK3 FDM printer. Measurements were performed in reflectance with the incident angle matching the angle of detection.

**Dynamic light scattering**

Temperature-dependent dynamic light scattering (DLS) measurements were performed with the Zetasizer Nano S (Malvern Panalytical) equipped with a laser of 633 nm wavelength and scattered light was detected in an angle of 173°. Measurements were conducted between 17°C and 60°C with steps of 0.2 °C. Three measurements with acquisition times of 60 s each were performed at each temperature step. The measurements were recorded in semi-macro cuvettes (polymethylmethacrylate, VWR) with CS microgel mass contents of 0.05 wt% in water. The hydrodynamic radii (z-average) were determined by cumulant analysis provided by the instrument software.

Angle-dependent DLS measurements were done on a 3D LS Spectrometer (LS Instruments, Switzerland) operated in 2D mode (pseudo cross-correlation) equipped with a HeNe laser (632.8 nm) as light source while the scattered light was detected with two avalanche photodetectors. A CS microgel dispersion, filtered through a 5 μm syringe filter (PTFE, Carl Roth) with a mass content of 0.01 wt% was prepared in a cylindrical glass cuvette (10 x 75 mm, borosilicate, Fisher scientific). Beforehand, the glass cuvette was treated with 2 vol% Hellmanex solution, followed by cleaning in an acetone fountain. Finally, samples were placed in a temperature-controlled decalin bath (JULABO CF31, PT100 close to sample position) and measured in angles between 30° and 140° in 5° steps with acquisition times of 60 s. The recorded data was analyzed by the CONTIN algorithm[45] as implemented in AfterALV v.1.0e (Dullware).

**Electrophoretic mobility determination**

The electrophoretic mobility of the CS microgels dispersed in water was determined at 20°C using a Litesizer 500 from Anton Paar.

**Transmission electron microscopy**

Transmission electron microscopy (TEM) was performed with a JEOL JEM-2100Plus TEM in bright-field mode. The acceleration voltage was set to 80 kV. Aqueous CS microgel dispersions were drop-casted on carbon-coated copper grids (200 mesh, Electron Microscopy Science) and dried for several hours at room temperature before investigation. TEM images were exported with the GMS 3 software from Gatan and the particle size was determined using ImageJ.[46]

## 3. Results and Discussion

**Characterization in the dilute regime**

We synthesized CS microgels that feature spherical silica nanoparticle cores and crosslinked PNIPAM shells. **Figure 1** summarizes the basic characterization in the dilute (non-interacting) regime. A schematic illustration of the microgel structure and the most relevant dimensions for this work are shown in **Figure 1a**. The SAXS profile in **Figure 1b** reveals the form factors measured at 20°C and 40 °C with contributions from the core in the high $q$ range and contributions from the shell in the mid to low $q$ range. Form factor analysis on the basis of a CS model with a homogeneous core and a shell with an exponentially decaying density profile (red line) provided a core radius of $R_{core}$ = 18 ± 2 nm and a total radius of the CS microgels of 138 ± 14 nm at 20 °C. More details on the form factor analysis including the application of other models can be found in the Supporting Information (**Table S3** and **Figure S1).** Due to the thermoresponsive behavior of PNIPAM based microgels, the CS microgels collapse above the volume phase transition temperature (VPTT). This is also reflected by the scattering profile recorded at 40 °C, i.e. well above the VPTT. In this state a simple CS model with homogeneous core and shell can be used to describe experimental scattering data. The respective fit to the data yields a total radius of 90 ± 8 nm. For more details on the fitting parameters see **Table S2** in the Supporting Information.

Temperature-dependent DLS measurements reveal the typical VPT behavior of the PNIPAM shell in aqueous dispersion (**Figure 1c**). In the swollen state (20 °C) the CS microgels possess a hydrodynamic radius of $R_h$ = 147 ± 3 nm. The radius continuously decreases with increasing temperature reaching a value of 97 ± 1 nm at 50 °C. The VPTT is approximately 35.0 °C. We want to note that we used 15 mol% of BIS in the synthesis (nominal) which explains the slightly higher temperature of the VPT as compared to 'classical' PNIPAM microgels. This is in agreement with findings from previous studies on comparable systems.[22, 47] Since the precision of the hydrodynamic dimensions in the swollen and collapsed state is of utmost importance for the further analysis in this work, we used angle-dependent DLS in addition to the fixed angle results. The good agreement between both sets of measurements can be seen in **Figure S2** in the Supporting Information.

**Figure 1d** shows a representative TEM image of the CS microgels where the cores and shells can be clearly distinguished due to their difference in contrast. The image supports that each microgel contains a single nanoparticle core with an average radius of $R_{core}$ = 17.5 ± 1.6 nm. SAXS measurements of the bare silica cores (**Figure S3**) are in good agreement with the results from TEM leading to an average radius of 17.8 ± 1.8 nm. A histogram from size analysis of a large number of microgels along with additional TEM images is provided in the Supporting Information (**Figure S4**). We found that less than 1% of the microgels feature more than one or no core. As expected, the microgel shells significantly shrink due to the sample preparation and the high vacuum conditions during TEM investigation. This is illustrated by the red circle in **Figure 1d** that highlights the swollen state dimensions determined by DLS.

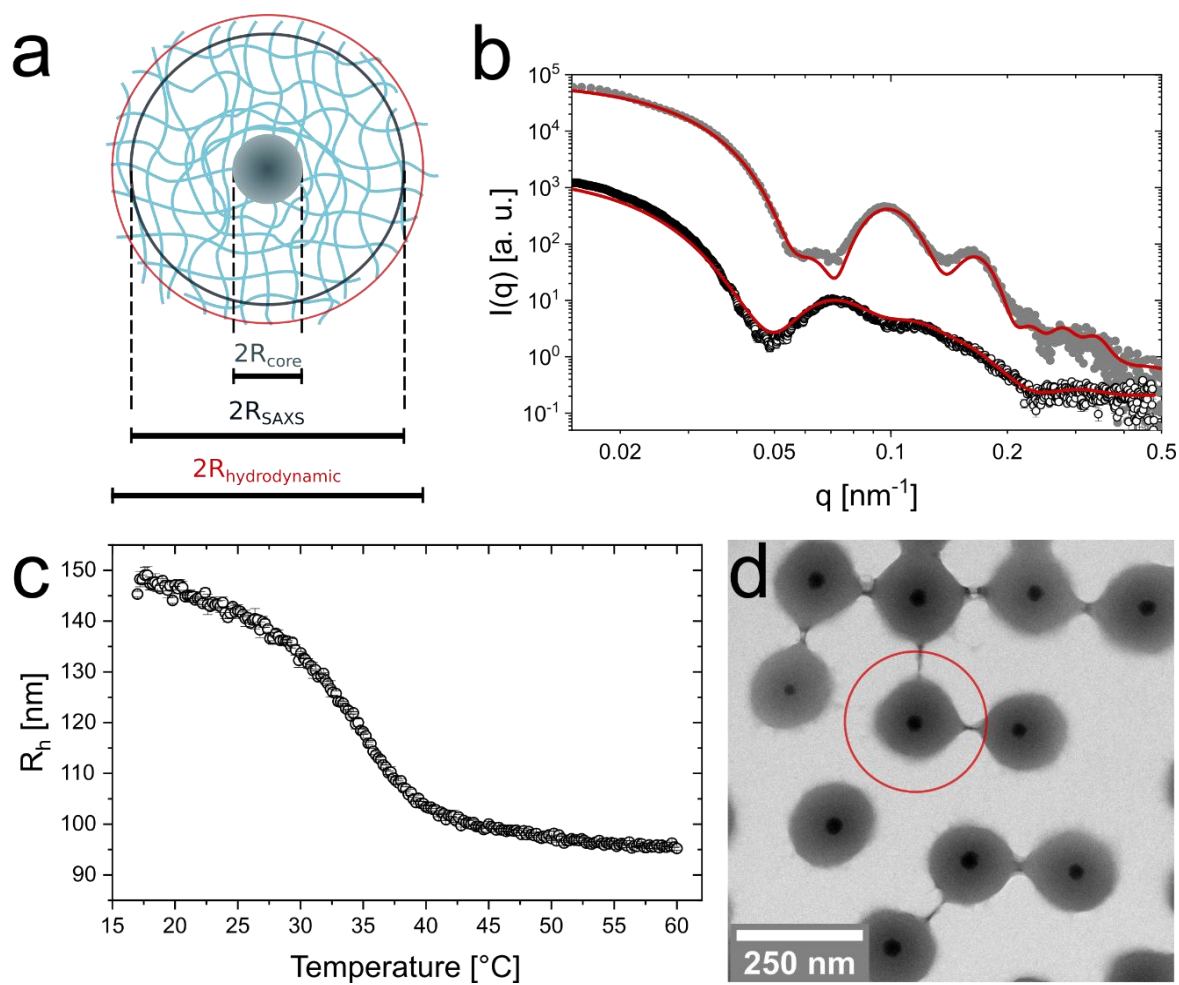

**Figure 1.** Characterization of the CS microgels in the dilute state. (a) Schematic illustration of a single CS microgel with the most relevant radii. $R_{core}$ is the radius of the core, $R_{SAXS}$ the total microgel radius from SAXS and $R_h$ is the hydrodynamic radius. (b) Synchrotron SAXS profile measured from a dilute dispersion (0.45 wt%) at 20 °C (black cirlces) and (1.25 wt%) at 40°C (grey dots). In case of the profile recorded at 20 °C, the solid line corresponds to a form factor model containing a homogeneous, isotropic core and a shell with an exponentially decaying density profile. A core-homogeneous-shell model was applied to fit the profile recorded at 40 °C. (c) Hydrodynamic radius as function of temperature (black circles) obtained from temperature-dependent DLS measurements (d) Representative TEM image. The red circle indicates the total dimensions obtained from DLS at 20°C.

**Fluid-solid transitions in dense packings**

Similar to hard spheres, microgels including our CS microgels are expected to crystallize at large enough packing fractions. Due to the total size of our CS microgels approaching visible wavelength scale total dimensions, absorbance spectroscopy is a convenient tool to monitor the phase behavior.[30, 47] **Figure 2a** shows Vis-NIR absorbance spectra of samples with 5.4 to 10.9 wt% CS microgels obtained directly after sample preparation (dashed lines) and after annealing at 50 °C followed by slow cooling to 20 °C (solid lines). More details on the annealing procedure and the parameters that we have identified for the best annealing conditions can be found in the Supporting Information (**Figure S5**). The annealing process reduces the linewidths of the narrow Bragg peaks, enhances their maximum intensity indicating a reduction in incoherent scattering. At the same time, only small spectral shifts are observed. For the lowest concentration (5.4 wt%) a Bragg peak is only observed prior to annealing. This can be attributed to a liquid-crystal coexistence phase and potentially local concentration differences right after sample preparation. During the annealing, the crystallites melt and the resulting fluid homogenizes resulting in a fluid phase that even remains after slow cooling to room temperature where the volume fraction is significantly increased. This behavior indicates that the volume fraction of this particular sample in the swollen state is close to the threshold volume fraction where crystallization occurs. In **Figure S6** in the Supporting Information, additional absorbance spectra recorded at different points in time can be found while **Figure S7** shows results from angle-dependent reflectance measurements. The reflectance data show the expected blue shift of the Bragg peak for increasing angle of incidence (with respect to the normal).

What happens during the heating and cooling cycles in our annealing protocol can be nicely followed by absorbance spectroscopy. Detailed information about the temperature cycling can be found in the Supporting Information (**Figure S8**). **Figure 2b** shows as an example the spectra recorded during the heating in 0.3 °C steps starting at 20 °C for the sample with 7.3 wt% CS microgels. During the first approximately 10 °C of heating the Bragg peak remains nearly unchanged in position, intensity and full width at half maximum (FWHM). When approaching a temperature of 30 °C and higher (inset in **Figure 2b**), we can see a decrease of the Bragg peak accompanied by a blue shift and broadening. These changes are related to a significant shrinking of the PNIPAM shell that reduces the microgel volume fraction $\phi$. This is related to the VPT of the microgels (see **Figure 1c**). At temperatures of 31.7 °C and higher, the Bragg peak cannot be observed anymore indicating that only a fluid phase is left. To follow this phase transition more closely, **Figure 2c** shows the evolution of the Bragg peak absorbance with temperature for a heating (red symbols) and cooling (blue symbols) series. The data for both, the heating and cooling cycle, almost perfectly overlap with very little hysteresis and slightly smaller inflection points for the cooling cycle (30.7 °C in contrast to 31.6 °C for the heating cycle). Surprisingly, the temperature window where the solid-fluid (fluid-solid) transition occurs is very narrow as indicated by the green vertical bar. Within approximately 0.6 °C the crystals melt during the heating cycle and reversibly recrystallize during the cooling cycle. It is important to note that data shown were not normalized or anyhow modified other than taking the maximum absorbance values at the respective position of the Bragg peak with only the incoherent background scattering being subtracted. This background subtraction is the reason why the plotted intensities remain nearly unchanged for temperatures outside the green-colored area. Following only the development of the coherent signal reveals the very sharp and reversible phase transition. The same heating/cooling study was also performed for the samples containing 9.1 and 10.9 wt% CS microgels. The respective data can be found in the Supporting Information (**Figure S9** and **Table S4**).

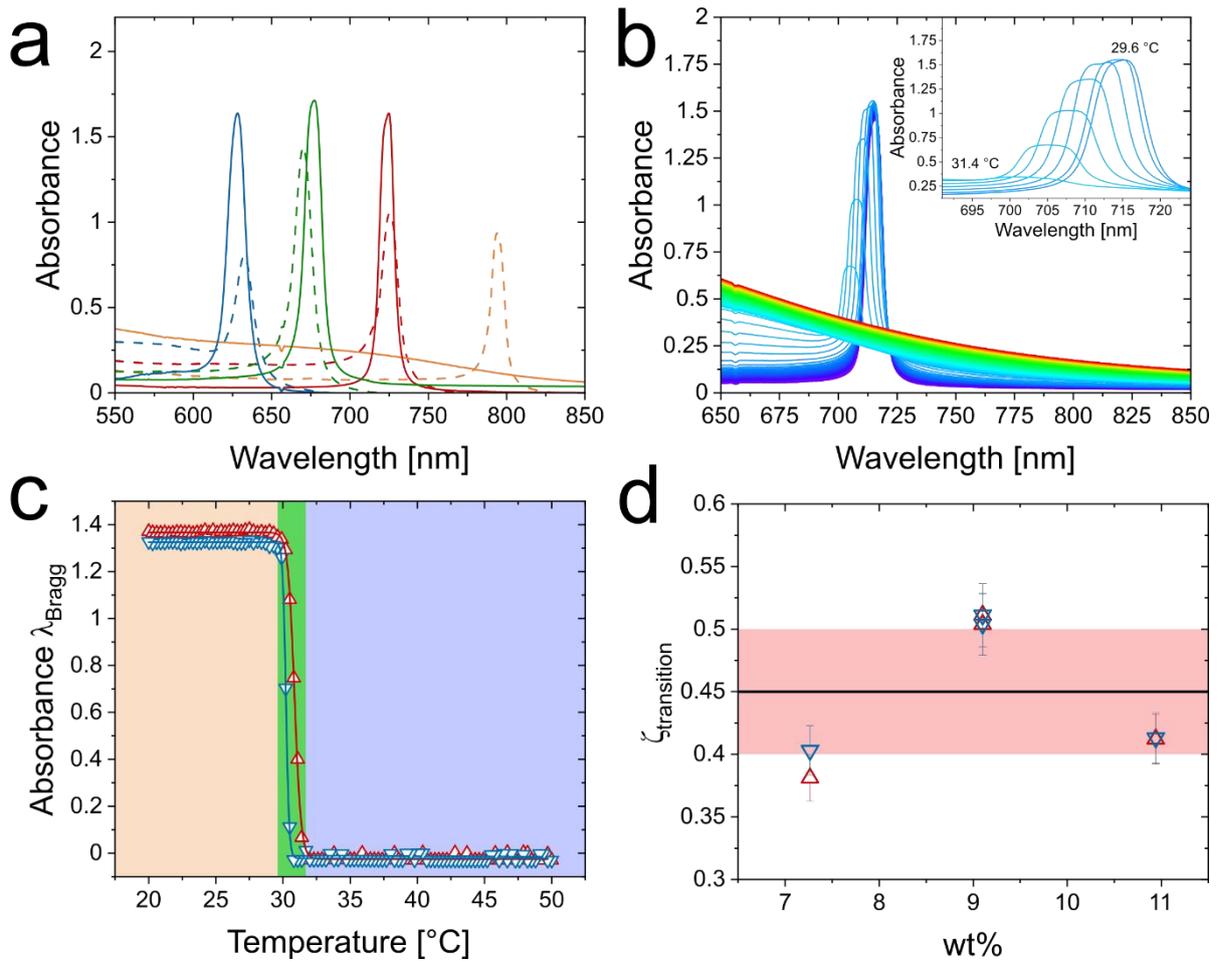

**Figure 2**. Phase behavior of CS microgels. (a) Vis-NIR absorbance spectra of CS microgel dispersion with mass contents of 5.4 wt% (orange), 7.3 wt% (red), 9.1 wt% (green) and 10.9 wt% (blue) after preparation (dashed lines) and after SAXS experiments (straight lines). (b) Temperature-dependent Vis-NIR absorbance spectra of 7.3 wt% CS microgel dispersion. Spectra were recorded from 20°C (violet) to 50°C (red) in 0.3 °C steps. The inset shows the temperature-dependent evolution of the Bragg peak at temperatures between 29.6°C and 31.4°C. (c) Temperature-dependent evolution of the intensity of the Bragg peak for the heating (red triangles) and cooling procedure (blue triangles). The green area indicates the temperature range where the Bragg peak disappears/reappears. (d) Threshold value of the generalized volume fraction where the transitions between fluid and crystalline phases occur in dependence of the mass contents of the respective samples. The horizontal, black line indicates the average volume fraction of 0.45 needed for crystallization and the red area is related to the respective standard deviation.

Similar to our previous works on gold-PNIPAM CS microgels,[22, 48] we can use the SAXS scattering contribution of the nanoparticle cores (here: silica) to determine the microgel number concentration, $N$. A detailed explanation for the data treatment and calculations is given in the Supporting Information (**Figures S10-12** and **Tables S5-8**). Having access to *N* allows us to determine the generalized volume fraction ζ that is based on the hydrodynamic microgel volume in the dilute, non-interacting regime available from our DLS data. **Figure S13** in the Supporting Information shows the evolution of ζ with temperature. We can now use the transition temperatures determined from the spectroscopic data, listed in **Table S4** of the Supporting Information, to determine the values of ζ where the solid-fluid (fluid-solid) transitions occur. **Figure 2d** shows the respective results for the heating and cooling cycles as a function of mass content of the samples. The data for the three different concentrations scatter around the average value of $\bar{\phi}_{transition}$ = 0.45 ± 0.05 (black line in **Figure 2d**) without a clear trend. The hysteresis between cooling (blue symbols) and heating (red symbols) is the largest for the sample at a concentration of 7.3 wt% and almost not observable for 9.1 and 10.9 wt%. We want to note that the volume fractions of the phase transition are small enough so that the generalized volume fraction should correspond to the real volume fraction.[48-49] Our mean value of 0.45 is slightly smaller than the freezing volume fraction of hard spheres (0.494).[50] We attribute this small deviation to an electrostatic contribution due to the anionic radical initiator used in the precipitation polymerization rendering the CS microgels charged.

**Structure in the fluid regime**

We performed synchrotron SAXS measurements on the dense samples at 40 °C, i.e. where the PNIPAM shells are collapsed. According to the analysis from absorbance spectroscopy, all samples should be in a fluid-like state at this temperature (see **Figure S9** in the Supporting Information) as Bragg peaks were not observed. **Figure 3a** shows a 2D detector image of a 9.1 wt% CS microgel dispersion recorded at 40 °C. A distinct liquid structure factor contribution, as well as some oscillations related to the form factor are visible in the isotropic scattering pattern. Additional detector images recorded from samples with different mass contents are provided in the Supporting Information (**Figure S14**). The results agree very well to the observation from absorbance spectroscopy. The radially averaged scattering profile is shown in **Figure 3b** (black circles). The profile shows a pronounced structure factor contribution in the range of low *q* (< 0.05 nm$^{-1}$). The mid to high *q* regime is dominated by the form factor of the CS microgels (see also **Figure 1b**). Scattering profiles for samples with 5.4 to 12 wt% CS microgel content are shown in **Figure S15** in the Supporting Information. The data reveal the expected changes in the structure factor, i.e. a shift of the first structure factor maximum to larger $q$ for increasing concentration. At the same time the form factor contribution remains unchanged. This implies that the microgel do not change due to osmotic deswelling in the collapsed state, at least in the range of concentrations studied. Consequently, the structure factor can be calculated directly from the scattering intensities of the dense samples ($I_{conc}(q)$) using the known form factor ($S(q) = I_{conc}(q)/P(q)$). Due to a small structure factor contribution in the experimental scattering profile of the dilute CS microgel dispersion (1.25 wt%) recorded at 40 °C (see **Figure S16**), we did not use the experimental data in the range of low $q$ (< 0.025 nm$^{-1}$) for this calculation. In the latter case, the theoretical form factor as shown by the fit in Figure 1b was used. **Figure 3b** shows the resulting structure factor (blue circles) for the 9.1 wt% sample. The vertical dashed line highlights the value in $q$ separating the range in $q$ where the form factor fit was used for the calculation ($q < 0.025$ nm$^{-1}$) and the range where the experimental data (1.25 wt%) were used ($q > 0.025$ nm$^{-1}$). More details on this data treatment are provided in the Supporting Information (**Figure S17**). The Percus-Yevick[51] (PY) hard sphere structure factor fit describes the determined structure factor almost perfectly.

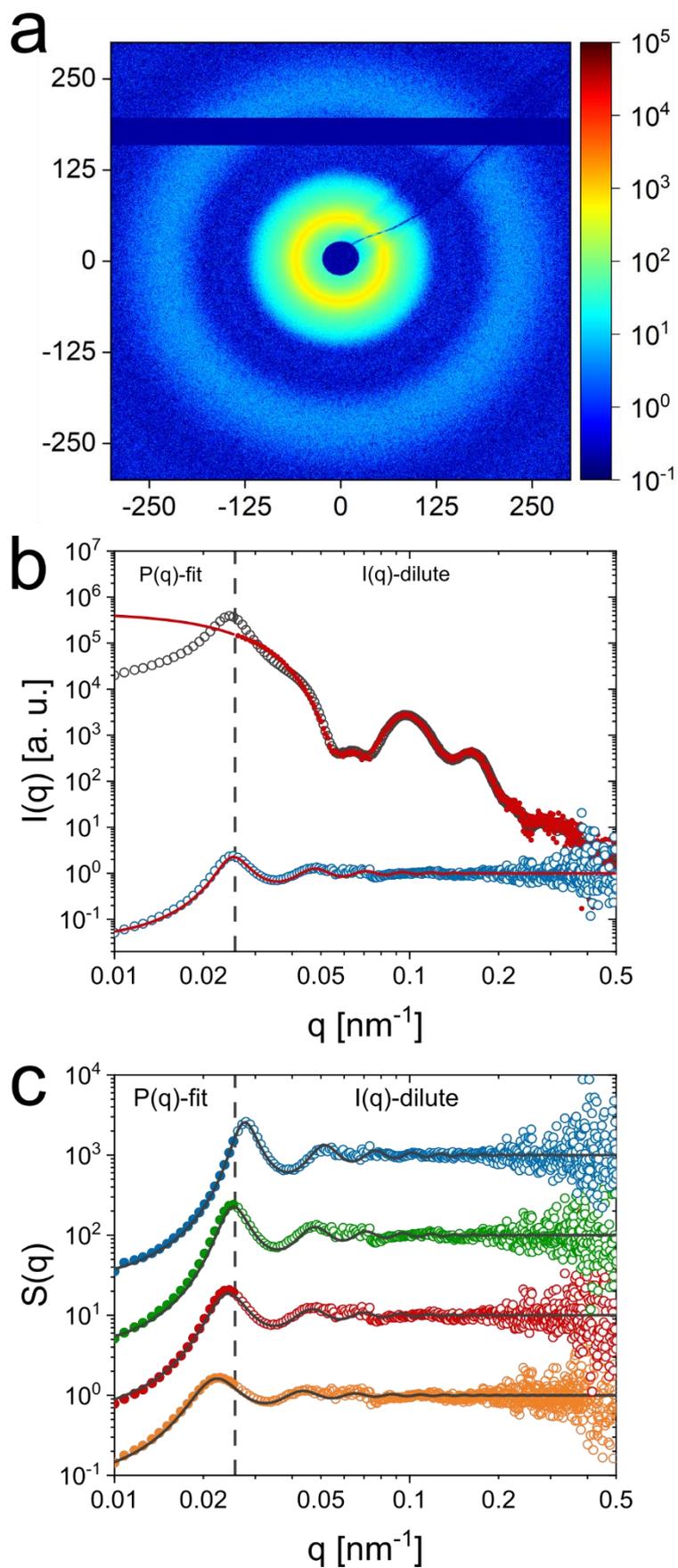

**Figure 3.** Synchrotron SAXS analysis of dense CS microgel dispersions, recorded above the VPTT at 40°C. Samples from 5.4 to 9.1 wt% were prepared in D$_2$O and the 12 wt%

dispersion in H₂O. (a) 2D detector image with a pronounced contribution of a liquid structure factor. (b) Structure factor (blue circles) extraction from the scattering profile of a 9.1 wt% CS microgel dispersion recorded at 40 °C (black circles). Solid red lines correspond to form and structure factor fits (Percus-Yevick) and red circles correspond to the (rescaled) experimental scattering profile of the dilute CS microgel dispersion at 40 °C. The dashed line highlights the threshold value where either the form factor fit ($q < 0.025$ nm$^{-1}$) or the experimental scattering signal of the dilute CS microgel dispersion ($q > 0.025$ nm$^{-1}$) was used for the structure factor calculation. (c) Extracted structure factors for CS microgel dispersions from scattering profiles recorded at 40 °C of 5.4 wt% (orange circles), 7.3 wt% (red circles), 9.1 wt% (green circles) and 12 wt% (blue circles). The solid black lines correspond to structure factor fits (Percus-Yevick). The structure factor profiles are shifted vertically for the sake of clarity.

**Figure 3c** compares the extracted structure factors for the samples in the range of 5.4 to 12 wt% with increasing concentration from bottom to top. All data can be nicely fitted by the PY structure factor model. We also observe the expected shift of the first structure factor maximum to larger $q$ for increasing concentration. At the same time, we observe a reduction in structure factor peak width. Results from the structure factor analysis are summarized in **Table S9** in the Supporting Information. The effective hard sphere volume fraction increases from 0.32 for 5.4 wt% to 0.46 for (12 wt%). At the same time the effective hard sphere radius decreases from 142 to 124 nm, values that significantly larger than the hydrodynamic radius in this collapsed state (40 °C) as determined from DLS. We attribute this decrease in hard sphere radius to an increasing electrostatic screening as the microgel and consequently the counterion concentration increases. electrostatic interactions that decrease when the counterion concentration is increased due to an increase in microgel concentration. **Figure S18** in the Supporting Information shows only the low $q$ region of the structure factors and reveals the good match between PY structure factor fit and experimental data towards $q = 0$ nm$^{-1}$. This underlines the hard sphere-like behavior of the CS microgels in the collapsed state. In swollen state experimental structure factors of microgels were shown to deviate from the hard sphere approximation which was related to the soft character of the interaction potential.[52]

**Structure in the solid regime (crystalline)**

The absorbance spectra of **Figure 2a** indicate that the crystallinity is significantly improved by the applied temperature annealing protocol. To verify this and to determine the crystal structure, we used synchrotron SAXS. **Figure 4a** shows the 2D detector image recorded for the 9.1 wt% sample at 20 °C, as a representative example. Pronounced and sharp Bragg peaks of at least 10 orders of diffraction are observed. Furthermore, the six-fold symmetry of the diffraction pattern is clearly visible indicating the alignment of hexagonal close-packed planes parallel to the capillary wall.[28] Apparently, SAXS studies on microgels and CS microgels in dense packings are scarce and from reported 2D diffraction patterns typically only up to four orders of diffraction are observed.[28, 48, 53] Although not directly comparable due to the different experimental smearing and detector resolution, analysis of colloidal crystals of microgels by small-angle neutron scattering (SANS) revealed diffraction with Bragg peaks of four diffraction orders.[47] The long-range order and superior crystallinity of our system can also be observed for the 7.3 and 10.9 wt% sample as shown in the Supporting Information (**Figure S19**).

**Figure 4b** shows the radially averaged data (black circles) corresponding to the detector image shown in **Figure 4a**. The structure factor with pronounced and well distinguishable Bragg peaks is clearly visible in the mid to low $q$ region ($q < 0.15$ nm$^{-1}$). In addition, the form factor minima related to the core ($q \approx 0.23$ nm$^{-1}$) and the microgel shell ($q < 0.1$ nm$^{-1}$) are clearly visible. Due to the distinct intensity minima, we can model the form factor even in the densely packed state where analysis is typically difficult/not possible due to changes in the microgel size and shape (osmotic deswelling and faceting)[40, 54-55] and the superposition with the structure factor. The red line in **Figure 4b** corresponds to our modeled form factor. Here, the CS structure of our microgels simplifies the modeling process, because the core contribution is well resolved and isolated from the shell as well as structure factor contribution. Consequently, also the particle number density, $N$, can be precisely determined through the form factor contribution of the core. The contribution of the shell can be modeled based on the distinct form factor minimum at $q \approx 0.05$ nm$^{-1}$. With the resulting form factor, $P(q)$, in the concentrated regime, we can calculate the structure factor by dividing the measured scattering intensity by $P(q)$ similar as has been presented before in the fluid regime.

The extracted structure factor is shown in **Figure 4b** (blue circles). In the high $q$ region $S(q)$ approaches the expected value of 1. This underlines that the modelled form factor describes the high $q$ region precisely. In the limit of low $q$, $S(q)$ approaches very low values close to zero. Extrapolation to $q = 0$ nm$^{-1}$ reveals a value $S(0) = 0.03$.

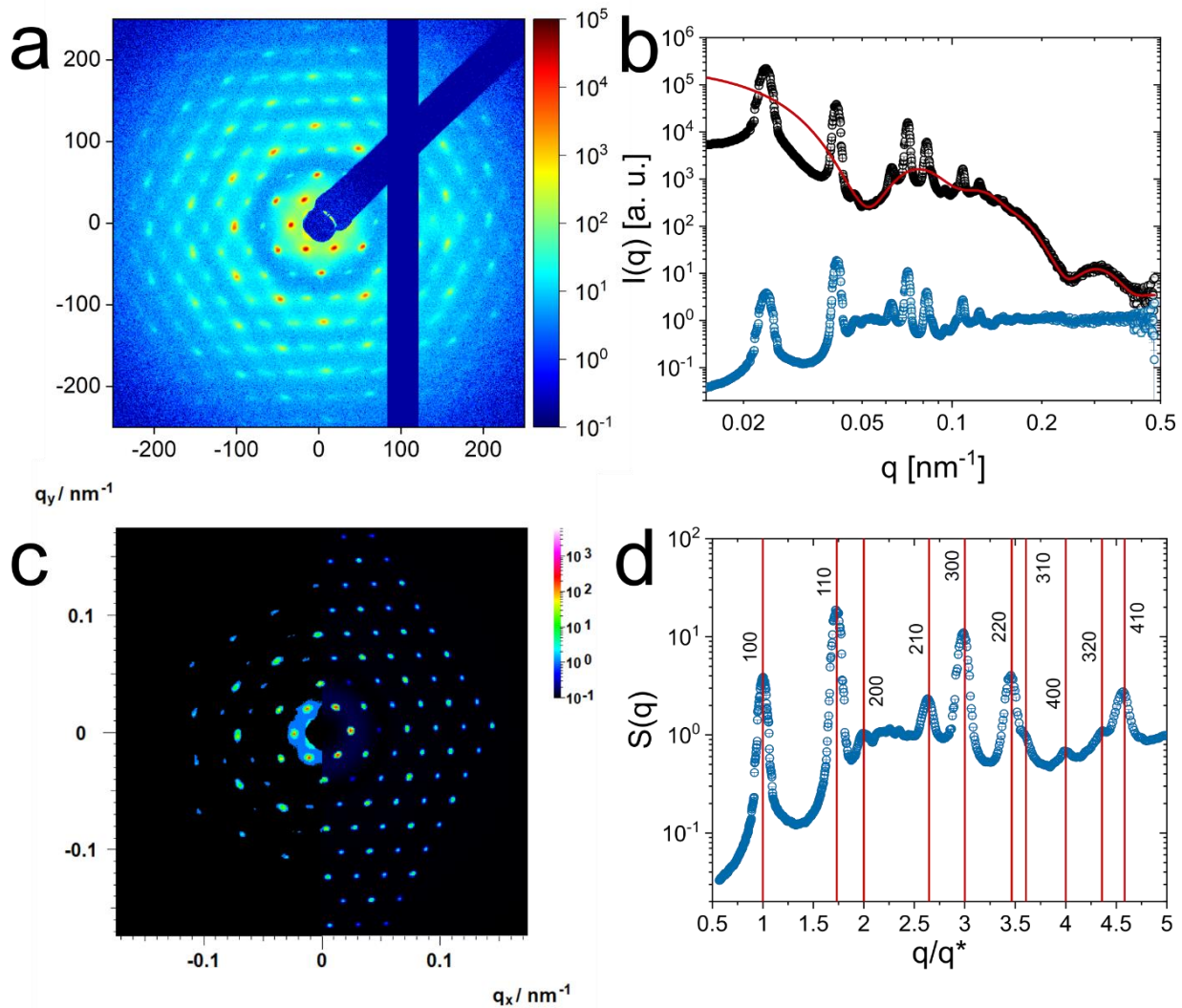

**Figure 4**. Synchrotron SAXS analysis of crystalline sample with 9.1 wt% CS microgels. (a) 2D detector image with pronounced Bragg peaks of several orders. (b) Corresponding radially averaged data (black circles). The red solid line corresponds to the modelled form factor. The blue circles correspond to the extracted structure factor. (c) Measured scattering data (2D) on the left and the corresponding hcp simulation on the right. (d) Structure factor from experiment with theoretical peak positions (vertical red lines) corresponding to a hcp crystal structure. The x-axis is normalized to the position of the first peak of the structure factor.

The large number of Bragg peaks allows us to simulate theoretical scattering patterns in direct comparison to the recorded 2D detector images. **Figure 4c** shows the perfect agreement with a hcp lattice. The Bragg peaks from simulation do not only match the experimental data in terms of peak positions and intensities but also the azimuthal and radial peak widths match closely. Detailed information on the simulation parameters and additional simulations performed on the 7.3 and 10.9 wt% samples are presented in **Table S10** and **Figure S20** in the Supporting Information.

The perfect agreement between measured peak positions and theoretical ones for a hcp lattice is also seen in **Figure 4d**. Here, the experimental structure factor is plotted with a normalized $q$ axis. Normalization was done based on the position of the first structure factor maximum, $q^*$. The vertical, red lines correspond to the calculated positions for a hcp lattice:

$$\frac{q}{q^*} = \frac{d_{hkl}}{d_{100}} = \frac{\sqrt{\frac{4}{3}(h^2+hk+k^2)+\frac{a^2}{c^2}l^2}}{\sqrt{\frac{4}{3}}} \quad (6)$$

Here, $d_{hkl}$ refers to the lattice spacing with Miller indices $h$, $k$ and $l$. In order to describe the dimensions of the unit cell $a$ refers to the lattice constant and $c$ to the height of the unit cell. The first Bragg peak to be expected is assigned to the $d_{100}$ lattice and for a closed packed structure of spheres the ratio between c and a is fixed to a value of $(8/3)^{1/2}$. The theoretical Bragg peaks which we could not identify in our scattering profile as well as a comparison to an fcc lattice are shown in **Figure S21** in the Supporting Information.

A similar analysis for samples in a range of concentrations from 5.4 to 10.9 wt% is presented in **Figure 5a**. Here the $q$ axis is again normalized by $q^*$. Only in case of the 5.4 wt% sample that revealed a mostly fluid character after long temperature annealing, the data were normalized by the second structure factor maximum (110 plane). The first structure factor maximum located at $q \approx 0.02$ nm$^{-1}$ is mostly attributed to a fluid like structure factor while the second one at $q \approx 0.03$ nm$^{-1}$ exhibits a small width and more distinctive shape and therefore is considered as a Bragg peak. We relate the Bragg peak to small residual crystallites. In **Figure S22** we show a fit of the fluid-like structure factor. From this, we conclude that the volume fraction of the CS microgels in the 5.4 wt% dispersion is very close to the freezing concentration. As **Figure 2c** indicates, only small change in temperature are required to induce the start of the crystallization or, vice-versa, melting.

For all samples with mass contents above 5.4 wt%, we find a nearly perfect agreement between the experimental and the theoretical positions of the Bragg peaks assigned to the hcp crystal structure. In **Figure 5b,** we show the linear relation of the Bragg peak position, $q_{hkl}$, and the *d*-spacing between lattice planes for hcp structures. The slopes of the linear fits provide the respective lattice constants, *a*, according to:

$$q_{hkl} = \frac{2\pi}{a}\left(\frac{4}{3}(h^2 + hk + k^2) + \frac{3}{8}l^2\right)^{\frac{1}{2}} \qquad (7)$$

**Table S11** in the Supporting Information lists the determined values of *a*. In addition, the lattice constant is accessible from the simulations of the 2D detector images like shown in **Figure 4c** as well from the positions of the Bragg peaks from absorbance spectroscopy, $\lambda_{\text{diff}}$:

$$a = \frac{\lambda_{\text{diff}}\sqrt{\frac{4}{3}(h^2 + hk + k^2) + \frac{3}{8}l^2}}{2\, n_{\text{crystal}}} \qquad (8)$$

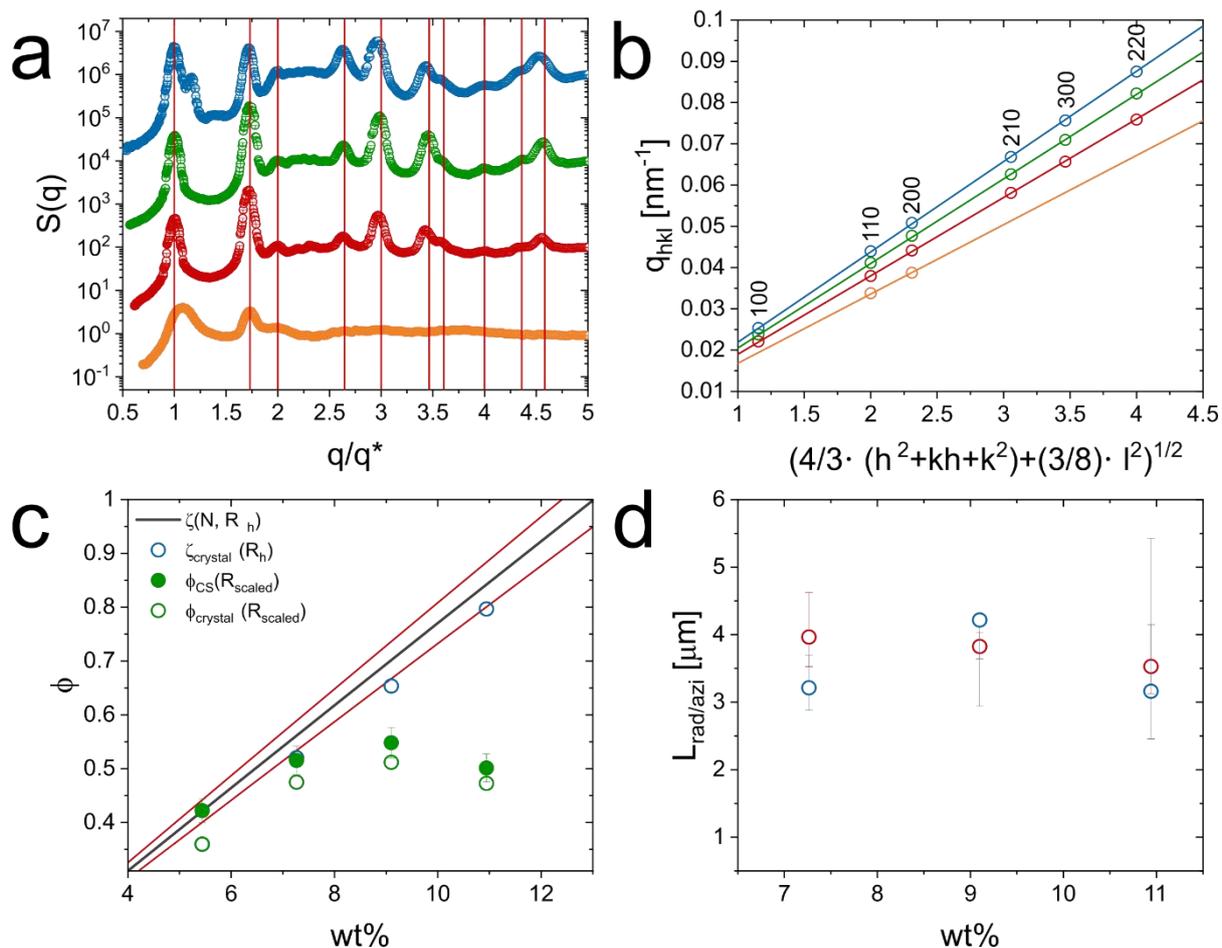

**Figure 5**. Extraction of the lattice constant *a* and the volume fraction $\phi$ of the CS microgels in the colloidal crystals. (a) S(q) of CS microgel dispersions with mass contents of 5.4 wt% (orange circles), 7.3 wt% (red circles), 9.1 wt% (green circles) and 10.9 wt% (blue circles). The straight vertical lines correspond to the theoretical Bragg peak positions assigned to the hcp crystal structure. The x-axis is normalized to the position of the respective first structure factor maximum except for the 6 wt% sample (orange circles) where the position was normalized by the position of the second peak. The profiles are offset along the ordinate for clarity. (b) Extraction of the lattice constant *a* via the linear dependence of the position of the Bragg peak on the *d*-spacing. (c) Volume fraction $\phi$ of the CS microgel in dependence of the mass content. The black line corresponds to the generalized volume fraction, based on particle number concentration and $R_h$ of the CS microgel. The red lines indicate the standard deviation. The blue circles and green dots are related to the volume fraction of the particles in the crystal based on the lattice constant and $R_h$ (blue) respectively $R_{h,scaled}$. Green circles indicate the volume fraction based on particle number concentration and $R_{h,scaled}$. (d) Radial and azimuthal sizes of the coherently scattering domains

For the hcp crystal structure, we assign the Bragg peak from the absorbance spectra to the 002 lattice. **Figure S23** in the Supporting Information shows the good agreement of the lattice constants, *a*, as obtained from analysis of SAXS data and absorbance spectra. The lattice constant decreases from 322 nm (7.3 wt% sample) to 284 nm (10.9 wt% sample). We can now calculate the volume fraction of the crystalline samples, $\phi_{\text{crystal}}$, based on the unit cell dimensions:

$$\phi_{\text{crystal}} = \frac{(3+3)\frac{4}{3}\pi(R_h)^3}{\frac{3\sqrt{3}\sqrt{\frac{8}{3}}a^3}{2}} = \frac{8\pi(R_h)^3}{3\sqrt{2}a^3} \tag{9}$$

**Figure 5c** compares the determined volume fractions based on the hydrodynamic radius in the dilute state, i.e. the generalized volume fraction, $\zeta$ (blue circles and black line) and using radii that we determined from the form factor analysis of the SAXS data directly in the densely packed state. The latter values allow us to rescale the hydrodynamic radius, $R_h$, using the obtained radii from SAXS in the dilute ($R_{\text{SAXS dil.}}$) and concentrated state ($R_{\text{SAXS}}$):

$$R_{h,\text{scaled}} = \frac{R_{\text{SAXS}}}{R_{\text{SAXS dil.}}} R_h \tag{10}$$

We see a good agreement between the volume fractions calculated based on $R_h$ with the number concentration $N$ (black line) and the volume fraction calculated with $R_h$ and the unit cell dimensions (blue circles) shown in **Figure 5c**. The red lines indicate the standard deviation of the calculated $\zeta$ based on the number concentration. The values of $\zeta$ obtained from the unit cell dimensions lie within the respective standard deviation. This underlines the reliability of the extraction of $\zeta$ via the number concentration obtained from the scattering of the core in SAXS.

The calculated values of the effective volume fractions are presented in **Figure 5c** as green symbols. The calculation is either based on the particle number concentration from SAXS measurements (filled, green symbols) or based on the lattice constants as obtained from SAXS measurements on crystalline samples (open, green circles). The data follow the same trend with slightly smaller volume fractions obtained from the crystal structure analysis. Again, the difference lies within the standard deviation with exception for the 5.4 wt% sample, where the lattice constant is related to small crystallites not representing the average bulk volume fraction of the dispersion. When reaching the threshold of crystallization (**Figure 2d**) we observe a significant deviation of the volume fraction from the generalized volume fraction. This is most pronounced for the 9.1 and 10.9 wt% samples where the volume fraction approaches values of approximately 0.52 which is far away from the closed packed volume fraction of 0.74 for hard spheres. This is in agreement to the values reported by Scotti based on comprehensive analysis of microgels in dense packing using SANS.[49]

In addition, we show the respective volume fractions based only on $R_{SAXS}$ without any normalization to the hydrodynamic radius $R_h$ in the Supporting Information (**Figure S24**). Due to the smaller total size of the CS microgels as observed by SAXS, the determined volume fractions are significantly smaller although still following the trend seen in **Figure 5c**.

To analyze the domain sizes of the crystalline samples, we use the Williamson-Hall analysis[56] that is based on the azimuthal and radial widths of the Bragg peaks as shown in **Figure S25** in the Supporting Information. This analysis revealed very low strains below 1% for our colloidal crystals and coherently scattering domain sizes between 3 and 4 µm (**Figure 5d**). These values are in good agreement with the slightly smaller domain sizes from analysis of the 2D detector images using Scatter. Based on the total size of our CS microgels we estimate that these domain volumes contain approximately 2000 microgels per single domain.

**Osmotic deswelling in the solid regime (crystalline)**

As reported in the literature, the concentration of counterions, which determines the osmotic stress of the suspension, grows with increasing microgel concentration for ionic microgels[36-37, 57] and also for microgels that are often considered neutral but typically carry some charges from the ionic initiator.[35, 58] As a consequence, when the bulk modulus of the microgels is lower than the external osmotic pressure the microgels start to deswell, even when they are not in direct contact.[35, 38, 49, 59] In our case, the CS microgels carry chargeable groups that are only attributed to the used anionic radical initiator. The electrophoretic mobility measured at 20 °C and dilute conditions is - 1.23 µmcm/Vs. Thus, our microgels can be considered as slightly charged. Due to the unique contrast situation of our CS microgels for SAXS, we can extract the form factor even in the regime of dense packings, i.e. where $S(q) \neq 1$. The measured SAXS profiles (symbols) and the modeled form factors (solid lines) are presented in **Figure 6a**. The high $q$ region can be very reliably described with the form factor model due to the contribution of the core. Since the rigid cores do not change in size, shape and its size polydispersity remains constant, the form factor contribution of the cores does not change. This is highlighted by the vertical solid line that marks the position of the form factor minimum due to the core contribution. In the mid and, in particular, low $q$ region where the structure factor contribution is strong, the fit and experimental data seem to deviate. This however is only caused by the absence of the structure factor in the model. The first minimum of the form factor contribution of the shell is still well-resolved and not hampered by the structure factor. Therefore, we get very good agreement between fit and data around the minimum (approximately 0.05 nm$^{-1}$). The parameters used for the form factor modeling are listed in **Table S12** in the Supporting Information. As the grey arrow indicates, the minimum associated to the microgel shell shifts towards lower $q$ with decreasing concentration. This means that the particle size decreases with increasing concentration. In other words, the microgel shells shrink as the particle number density increases.

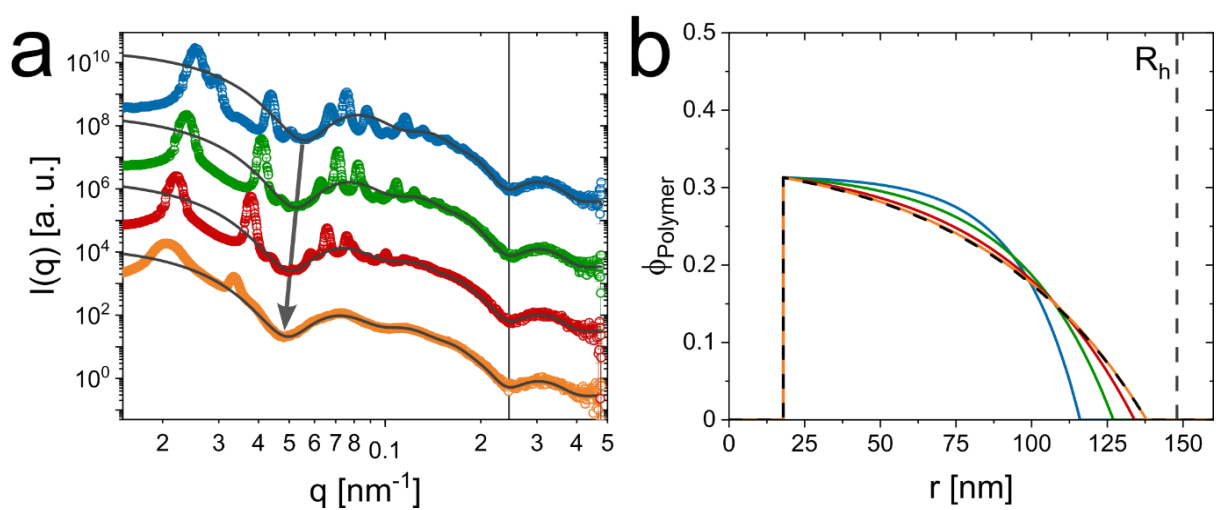

**Figure 6**. Osmotic deswelling of the CS microgels in dense packings. (a) Scattering profiles from samples with mass contents of 5.4 wt% (orange circles), 7.3 wt% (red circles), 9.1 wt% (green circles) and 10.9 wt% (blue circles). The grey lines correspond to the respective form factors and the grey arrow illustrates the shift of the form factor minimum related to the shell, while the black, vertical line indicates the fixed position of the form factor minimum related to the core. The profiles are offset along the ordinate for clarity. (b) Radial density profiles of the polymer volume fraction at the respective mass contents (similar color coding to a). The black dashed profile corresponds to the CS microgels in the dilute state and the black dashed, vertical line highlights the hydrodynamic radius.

**Figure 6b** shows radial polymer density profiles as the result from the form factor analysis. The respective polymer volume fractions are extracted from a radial density profile of the scattering length density shown in **Figure S26** in the Supporting Information. Here we also present the constant integrals of the radial density profiles. The rigid silica cores have a mean radius of 18 nm. Consequently, in **Figure 6b** the polymer volume fraction is zero for $r < 18$ nm. Within the shell region ($r > 18$ nm) the polymer volume fraction decreases from very similar values of 0.31 independent on the mass content of the samples to zero following the exponential decay of the form factor model. With an increase in mass content, we can detect a decrease of $R_{SAXS}$ from 138 nm to 116 nm and a more pronounced decline in the exponential decay. As we could not detect an increase in the polydispersity of the CS microgels and due to the dense packing of the CS microgels inside of the colloidal crystal they undergo an isotropic osmotic deswelling.[40, 49, 60] This leads to the decrease in radius and the more pronounced decline of the polymer volume fraction as described above. From literature it is known that the internal microgel architecture including the nominal crosslinker content dictates whether isotropic deswelling or faceting of the microgels is observed.[13] For microgels with high nominal crosslinker contents (≥ 5 mol% BIS), like in our case, it is known that isotropic deswelling occurs before faceting.[20, 49] A similar behavior was found for hollow microgels that showed isotropic osmotic deswelling when the particle concentration in the dispersion was increased. In this case it was attributed that deswelling is favored over faceting due to the presence of the cavity inside the microgel which allows for rearrangement of the polymer chains within the microgel.[54, 61] Ultra low crosslinked microgels, expressing pronounced soft behavior, show faceting when they are dispersed in a crowded environment, rather than undergoing isotropic deswelling before microgels make contacts.[49, 62]

## 4. Conclusion

Core-shell microgels with polymeric shells that are significantly larger than the incompressible core are ideal candidates to study fluid-solid transitions by small-angle scattering without the need of selective deuteration and contrast variation.[20, 34, 49] We have shown that SAXS is particularly powerful given that the core as well as the shell provide sufficient contrast with respect to the solvent environment. In our case this was realized by silica-PNIPAM core-shell microgels with shell-to-core size ratios on the order of 8 (swollen state). The form factor contribution of the silica cores in the range of large $q$ is well separated from the scattering of the shell and also from structure factor contributions in dense packings. Due to the overall microgel dimensions approaching the wavelength of visible light, crystallization of the microgels can be easily followed optically due to the photonic properties. We used absorbance spectroscopy as an efficient and easily available method to study solid-fluid/fluid-solid transitions induced by heating/cooling of the samples. Analysis of the data revealed millikelvin temperature windows where the phase transitions occur with almost no hysteresis between heating and cooling. Freezing point is registered at a volume fraction of 0.45 ± 0.05. The deviation from the value of hard spheres was explained by electrostatic interactions. The structures of the fluid and solid phases were studied by synchrotron SAXS. Structure factors extracted in the fluid regime showed perfect agreement to the Percus-Yevick model for hard spheres. In addition, we did not observe any osmotic deswelling in dense packings of collapsed microgels. In contrast, the samples in the solid regime showed sharp Bragg peaks and many different diffraction orders upon very slow temperature annealing of the samples. Analysis of the scattering data revealed hcp structures and domain sizes in the micrometer range. From the scattering profiles of the crystalline samples, we could also directly determine the form factor of the CS microgels. Data analysis revealed osmotic deswelling with an extend that increases with increasing concentration.

The unique combination of our silica-PNIPAM core-shell microgels and properly chosen shell-to-core size ratio with synchrotron SAXS is ideally suited to determine form as well as structure factors even in packings exceeding the hard sphere limit. Due to the great resolution of SAXS and the extremely short acquisition times, in situ studies on phase transitions over a broad range of packing densities will be possible in future works.

**Conflicts of interest**

There are no conflicts of interest to declare.

**Acknowledgements**

The authors acknowledge the German Research Foundation (DFG) and the state of NRW for funding the cryo-TEM (INST 208/749-1 FUGG) and Marius Otten from Heinrich-Heine-University Düsseldorf for his assistance with the operation and image recording. The authors thank the Center for Structural Studies (CSS) that is funded by the DFG (Grant numbers 417919780 and INST 208/761-1 FUGG) for access to the SAXS instrument. A. S. thank the DFG for financial support of project A3 (Project No. 191948804) within the SFB 985 – Functional Microgels and Microgel Systems. The authors acknowledge Déborah Feller for her assistance with the operation of the angle-dependent dynamic light scattering setup. A. V. Petrunin, T. Höfken, and P. Mota-Santiago helped with the SAXS measurements. The SAXS measurements were performed on the CoSAXS beamline at the MAX IV laboratory (Lund, Sweden) under the proposals 20200777 and 20220526. The Research conducted at MAX IV, a Swedish national user facility, is supported by the Swedish Research council under contract 2018-07152, the Swedish Governmental Agency for Innovation Systems under contract 2018-04969, and Formas under contract 2019-02496. This work benefited from the use of the SasView application, originally developed under NSF award DMR-0520547. SasView contains code developed with funding from the European Union's Horizon 2020 research and innovation programme under the SINE2020 project, grant agreement No 654000.

**TOC:**

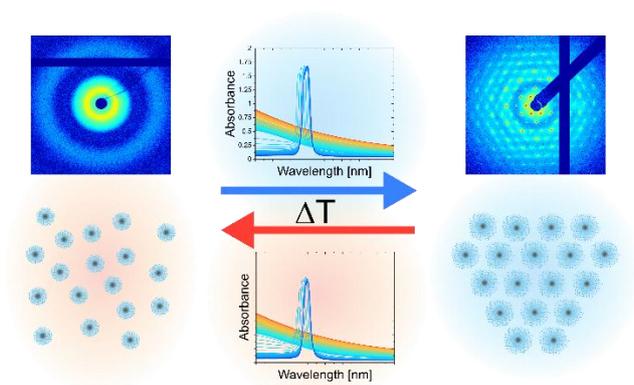

# Supporting Information

**Fluid-solid transitions in photonic crystals of soft, thermoresponsive microgels**

M. Hildebrandt,[1] D. Pham Thuy,[1] J. Kippenberger,[1] T. L. Wigger,[1] J. E. Houston,[2] A. Scotti,[3] and M. Karg[1]*

[1]Institut für Physikalische Chemie I: Kolloide und Nanooptik, Heinrich-Heine-Universität Düsseldorf, Universitätsstraße 1, D-40225 Düsseldorf, Germany

E-Mail: karg@hhu.de

[2]European Spallation Source ERIC, Box 176, SE-221 00 Lund, Sweden

[3]Institute of Physical Chemistry, RWTH Aachen University, Landoltweg 2, 52056 Aachen, Germany

**List of samples**

Core-shell (CS) microgel dispersions were prepared with various mass contents and filled into different capillaries depending on the type of measurements. Detailed information about the sample composition and the used capillaries are listed in **Table 1**. In order to maintain comparability between the samples prepared in $H_2O$ and $D_2O$ we used the same volume of solvent for dispersions made with the same mass of CS microgels. Thus, the dispersions possess the same particle number concentration in $H_2O$ and $D_2O$. As an example, the 10.9 wt% sample prepared in $D_2O$ and the 12 wt% sample prepared in $H_2O$ have the same particle number concentration.

**Table 1.** List of the CS microgel dispersions prepared for the different experiments.

| Mass content [wt%] | Dispersant | Capillary | Experiment |
|---:|:---:|:---:|:---:|
| 0.45 | $D_2O$ | rectangular | SAXS (20 °C) |
| 5.4 | $D_2O$ | rectangular | SAXS (20 °C) / absorbance (various *T*) |
| 7.3 | $D_2O$ | rectangular | SAXS (20 °C) / absorbance (various *T*) |
| 9.1 | $D_2O$ | rectangular | SAXS (20 °C) / absorbance (various *T*) |
| 10.9 | $D_2O$ | rectangular | SAXS (20 °C) / absorbance (various *T*) |

| 1.25 | H₂O | round | SAXS (40 °C) |
| 5.4 | D₂O | round | SAXS (40 °C) |
| 7.3 | D₂O | round | SAXS (40 °C) |
| 9.1 | D₂O | round | SAXS (40 °C) |
| 12 | H₂O | round | SAXS (40 °C) |
| 10 | H₂O | round | SAXS (determination of $N$) |

**Form factor analysis**

We measured SAXS from dilute and concentrated core-shell (CS) microgel dispersions and radially averaged the scattering data to obtain the scattering intensity $I$, as function of the magnitude of the scattering vector $q$.

$$|\vec{q}| = q = \frac{4\pi}{\lambda} \sin\frac{\theta}{2} \quad (S1)$$

The scattering angle is given by $\theta$ and $\lambda$ is the wavelength of the X-rays.

In general, for the investigated $q$-range, the scattering intensity can be described by the following equation:

$$I(q) = N V_{\text{particle}}^2 \Delta SLD^2 P(q) S(q) + I_B \quad (S2)$$

With $N$ corresponding to the particle number density, $V_{\text{particle}}$ the volume of the scattering object and $\Delta SLD$ the difference in scattering length density (SLD) between the scattering object and the respective solvent. $P(q)$ is the form factor of the scattering object, $S(q)$ is the structure factor and $I_B$ describes additional incoherent background contributions. For dilute dispersions, the structure factor $S(q) \approx 1$.

The form factor modeling was performed with the SasView software[1]

To describe the form factor of our CS microgels, we used a CS model:[2]

$$I(q) = \frac{scale}{V_{\text{particle}}} \Bigg[ 3V_{core} SLD_{core} \frac{\sin(qR) - qR_{core}\cos(qR_{core})}{(qR_{core})^3} +$$

$$3V_{particle} SLD_{shell} \frac{\sin(qR_{\text{particle}}) - qR_{\text{particle}}\cos(qR_{\text{particle}})}{(qR_{\text{particle}})^3} -$$

$$3V_{\text{particle}} SLD_{solvent} \frac{\sin(qR_{\text{particle}}) - qR_{\text{particle}}\cos(qR_{\text{particle}})}{(qR_{\text{particle}})^3} \Bigg]^2 + I_B \quad (S3)$$

With *scale* as scaling factor (corresponding to the volume fraction if measured in absolute units), $V$ representing the volume of the respective scattering object and $R_{\text{core}}$ and $R_{particle}$ corresponding to the total radius of the core and particle. Here, $R_{particle}$ is the sum of $R_{core}$ and the thickness of the shell $\Delta t_{shell}$. Due to their homogeneous structure, the *SLD* of the solvent and the core is kept constant. In case of the CS microgels in their collapsed state, we used a constant *SLD* for the shell. For the CS microgels in their swollen state, we used an exponential decay in the *SLD* profile of the shell.

The exponentially decaying contrast of the shell is described by:

$$SLD_{shell}(R) = B\, exp\left(\frac{A(R-R_{core})}{\Delta t_{shell}}\right) + C \tag{S4}$$

With $\Delta t_{shell}$ being the thickness of the shell, and *A* describing the decay. *B* and *C* are defined as:

$$B = \frac{\rho_{out}-\rho_{in}}{e^A-1} \tag{S5}$$

$$C = \frac{\rho_{in}e^A-\rho_{out}}{e^A-1} \tag{S6}$$

Here, $\rho_{in}$ and $\rho_{out}$ are related to the SLD of the shell at $R_{in} = R_{core}$ and $R_{out} = R_{particle}$

The exponential SLD profile is described by the following equation:

$$f_{shell} = 3BV(R_{shell})e^A h(\alpha_{out}, \beta_{out}) - 3BV(R_{core})h(\alpha_{out}, \beta_{out}) +$$
$$3CV(R_{shell})\frac{\sin(qR_{shell})-qR_{shell}\cos(qR_{shell})}{(qR_{shell})^3} - 3CV(R_{core})\frac{\sin(qR_{core})-qR_{core}\cos(qR_{core})}{(qR_{core})^3} \tag{S7}$$

where

$$\alpha_{in} = A\frac{R_{core}}{\Delta t_{shell}} \tag{S8}$$

$$\alpha_{out} = A\frac{R_{shell}}{\Delta t_{shell}} \tag{S9}$$

$$\beta_{in} = qR_{core} \tag{S10}$$

$$\beta_{out} = qR_{shell} \tag{S11}$$

and

$$h(x,y) = \frac{x\sin(y)-y\cos(y)}{(x^2+y^2)y} - \frac{(x^2-y^2)\sin(y)-2xy\cos(y)}{(x^2+y^2)^2 y} \tag{S12}$$

The polydispersity of the core and the shell of our CS microgels are included using a Gaussian distribution of the respective radii. Here, $\langle r \rangle$ is related to the average particle radius, and $\sigma_{poly}$ describes the relative size polydispersity.

$$D(R, \langle R \rangle, \sigma_{poly}) = \frac{1}{\sqrt{2\pi\sigma_{poly}^2}} exp\left(-\frac{(R-\langle R \rangle)^2}{2\sigma_{poly}^2}\right) \tag{S13}$$

Due to the distinct difference in core and shell size, we can also fit the core contribution only. To describe the scattering intensity of the silica cores, we applied a simple polydisperse sphere model:

$$I(q) = \frac{scale}{V_{particle}}\left[3V(R)\Delta SLD \frac{\sin(qR)-qR\cos(qR)}{(qR)^3}\right]^2 + I_B \tag{S14}$$

The parameters used for the fits are listed in **Table S2**.

Table S2. Parameters applied to fit the scattering profiles of the dilute CS microgel dispersions in their swollen (T = 20 °C) and collapsed state (T = 40 °C).

| Parameters | Core-exponential-shell (swollen state) | Core-homogeneous-shell (collapsed state) |
|---|---|---|
| *scale* | 0.029 | 2.733 |
| $I_B$ [a. u.] | 0.2 | 0.003 |
| $R_{core}$ [nm] | 18 | 18 |
| $\Delta t_{shell}$ [nm] | 120 | 72 |
| $SLD_{core}$ [$10^{-6}$ Å$^2$] | 17.75 | 17.75 |
| $SLD_{shell, in}$ [$10^{-6}$ Å$^2$] | 9.89 | 10.30 |
| $SLD_{shell, out}$ [$10^{-6}$ Å$^2$] | 9.43 | - |
| $SLD_{solvent}$ [$10^{-6}$ Å$^2$] | 9.43 | 9.43 |
| $\sigma_{core}$ | 0.1 | 0.1 |
| $\sigma_{shell}$ | 0.1 | 0.08 |
| A | 2.2 | - |

We began the fitting procedure using fixed values for the background ($I_B$), $R_{core}$, $\sigma_{core}$, $SLD_{solvent}$, and $SLD_{core}$. Prior to the first fitting-steps, reasonable values for the $\Delta t_{shell}$, $SLD_{shell}$ and $\sigma_{shell}$ were assumed to get a starting point for the fitting-procedure. Next, the parameters were fitted in the following order: *scale*, $SLD_{shell}$, $\Delta t_{shell}$, A (decay constant), $\sigma_{shell}$. The values obtained from each fitting-step were applied for the following step. In each step, only the respective parameter was set free to change, while the other parameters were kept constant. As the model consists of many parameters, setting all parameters free resulted in a fit no longer describing the data sufficiently well when using SasView. Therefore the procedure was repeated until the fit described the experimental data sufficiently well.

Prior to the synchrotron SAXS experiments, $R_{core}$ and $\sigma_{core}$ were determined from inhouse SAXS measurements presented in **Figure S6**. The q-range in the experiment was limited in the low-$q$ regime and we were only able to fit the scattering of the core. Due to this, the values were already known and could kept constant in the fitting-procedure regarding the CS microgels.

We also applied other form factor models including the core-shell microgel model that is based on the widely accepted fuzzy-sphere model to describe the inhomogeneous microgel.[3] **Figure S1** compares the results for a simple CS model (homogeneous shell) in orange, the core-shell microgel model that is described by a box like profile for the core and the inner homogeneous region of the microgel shell with respective constant SLDs followed by an error function like decay similar to the fuzzy sphere model in green[4] and the previously discussed CS model with exponentially decaying shell in red. These additional form factor fits were performed with the SASfit software.[5]

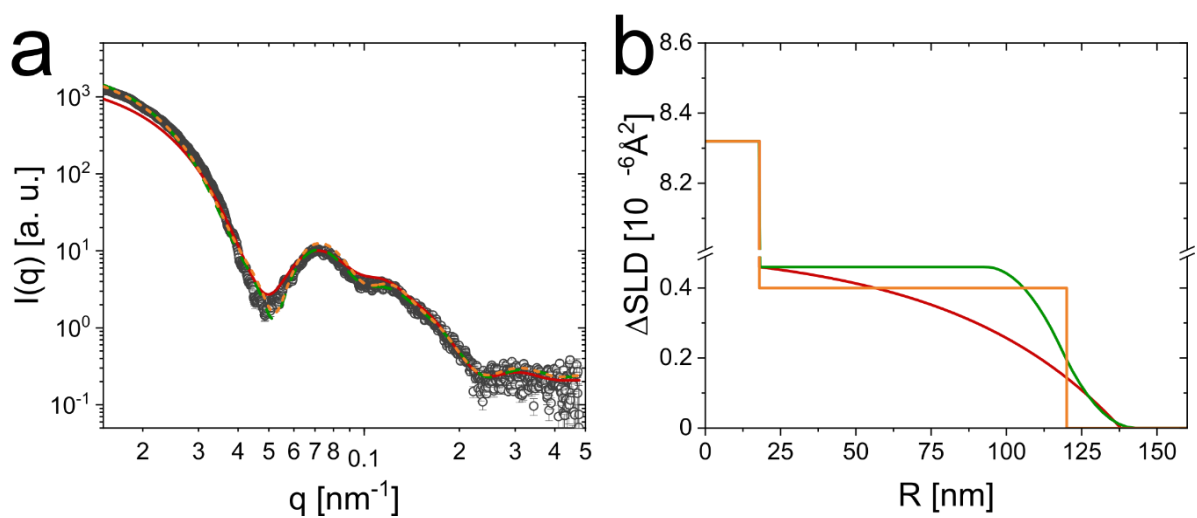

**Figure S1**: Comparison of different form factor models. (a) SAXS profile recorded from a dilute CS microgel dispersion with solid lines corresponding to fits using a core-shell model with an exponentially decaying shell (red), the core-shell microgel model (green, dashed line) and a core-shell model possessing a homogeneous shell (orange, dashed line). (b) Radial density profiles from the form factor fits with the same color coding as in (a).

**Figure S1a** shows that all three applied CS models can describe the SAXS profile of our CS microgels quite well. The parameters used to fit the scattering intensities are listed in **Table S3**. The radial density profiles extracted from the respective CS models are shown in **Figure S1b**. All models exhibit a high ΔSLD value related to the core until a radius of 18 nm. From this point on, we can see an exponential decay for the core-exponential-shell model (red) and a constant ΔSLD for the core-shell model (orange). The core-shell microgel model (green) exhibits the box like profile with a constant ΔSLD followed by a decay ascribed to the fuzzy surface of a microgel. The core-exponential-shell and the core-shell microgel model result in similar total radii of 138 ± 14 nm and 143 ± 27 nm, while the core-shell model yields a much smaller radius with just 120 nm. In order to obtain reasonable fits with the core-shell microgel model, a polydispersity as large as 20% was imposed on the homogeneous box profile, which is very unrealistic according to our results from dynamic light scattering, where the polydispersity was below 10%. We conclude that the core-exponential-shell model leads to more realistic results for the size and morphology of the CS microgels. Due to the lower number of parameters, we decided to use the core-exponential-shell model in the following of this work.

**Table S3**. Fit parameters obtained from different form factor models.

| Parameters | Core-exp.-shell | Core-shell microgel | Core-shell |
|---|---|---|---|
| $R_{core}$ [nm] | 18 | 18 | 18 |
| $\Delta t_{shell}$ [nm] | 120 | 75 | 102 |
| $sigma_{out}$ [nm] | - | 25 | - |
| $SLD_{core}$ [$10^{-6}$ Å$^2$] | 17.75 | 17.75 | 17.75 |
| $SLD_{shell,}$ [$10^{-6}$ Å$^2$] | 9.89 | 9.89 | 9.83 |
| $SLD_{solvent}$ [$10^{-6}$ Å$^2$] | 9.43 | 9.43 | 9.43 |
| $\sigma_{core}$ | 0.1 | 0 | 0 |
| $\sigma_{shell}$ | 0.1 | 0.2 | 0.15 |
| $A$ | 2.2 | - | - |

## Dynamic light scattering

In addition to the temperature dependent DLS measurements shown in **Figure S2a**, we performed angle-dependent DLS measurements below and above the volume phase transition temperature (VPTT) at 20°C and 45°C, respectively. The decay constants, $\Gamma$, were computed from the normalized field-time autocorrelation functions using the CONTIN algorithm[6] via the AfterALV software (v1.06d, Dullware, Amsterdam, The Netherlands).

Plotting the values of $\Gamma$ against $q^2$ (**Figure S2b**) we can extract the translational diffusion coefficient, $D_T$, with high precision:

$$D_T = \frac{\bar{\Gamma}}{q^2} \tag{S15}$$

The Stokes-Einstein equation can then be applied to determine the hydrodynamic radius $R_h$:

$$D_T = \frac{kT}{6\pi\eta R_h} \tag{S16}$$

Here, $k$ is the Boltzmann constant, $T$ is the absolute temperature and $\eta$ is the viscosity of the solvent.

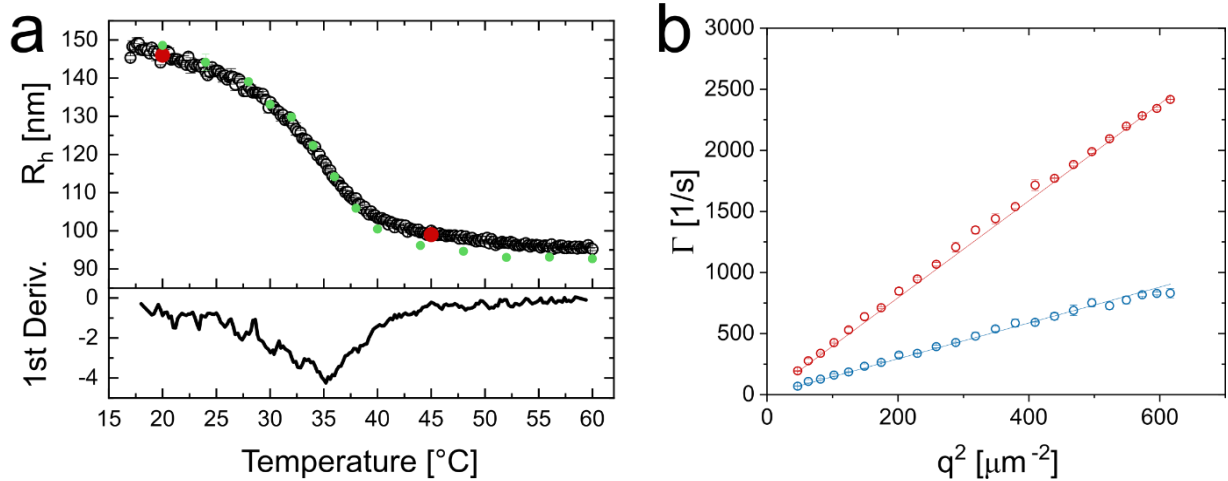

**Figure S2**. (a) Hydrodynamic radius as function of temperature obtained from fixed angle measurements in $H_2O$ (black circles, Zetasizer), $D_2O$ (green circles, Zetasizer) and angle-dependent DLS (red circles, $H_2O$). (b) Decay constants from angle-dependent DLS as function of $q^2$. The blue circles correspond to the data recorded at 20 °C and the red circles were recorded at 45 °C. The straight lines correspond to linear fits to the data.

Translational diffusion coefficients of $1.47 \cdot 10^{-12}$ m²/s and $3.97 \cdot 10^{-12}$ m²/s were extracted from the linear fits and correspond to $R_h$ of 146 ± 1 nm at 20°C and 99 ± 1 nm at 45 °C presented as red circles in **Figure S2a**. The obtained values form angle-dependent DLS match well to the results from the temperature dependent measurements with a $R_h$ of 147 ± 3 nm at 20 °C and 100 ± 1 nm at 45 °C. In addition, we extracted a VPTT of 35.1 °C for the CS microgels based on the 1st derivative of the temperature dependent DLS data. To investigate isotope effects on the thermoresponsive properties of the CS microgels we performed temperature dependent DLS measurements in $D_2O$ as solvent. The $R_h$ is shown as function of the temperature as green circles in **Figure S2a**. A deviation from the hydrodynamic radii recorded in $H_2O$ is only visible for temperatures above 38 °C as the particles exhibit a $R_h$ of 100 ± 1 nm in $D_2O$ and 103 ± 1 nm in $H_2O$ at a temperature of 40 °C. Up to the temperature of 38 °C, the hydrodynamic radius determined in $H_2O$ describes the size of the CS microgels sufficiently well, independent of the here used solvents.

**SAXS investigation of the silica cores**

The scattering profile of the $SiO_2$ cores prior to the encapsulation in the PNIPAM shell is presented in **Figure S3** and shows multiple form factor oscillations between $q = 0.15 - 1.0$ nm⁻¹. We fitted the scattering data with a homogeneous sphere model (red solid line) yielding a radius of 18 ± 2 nm with a polydispersity of 10%.

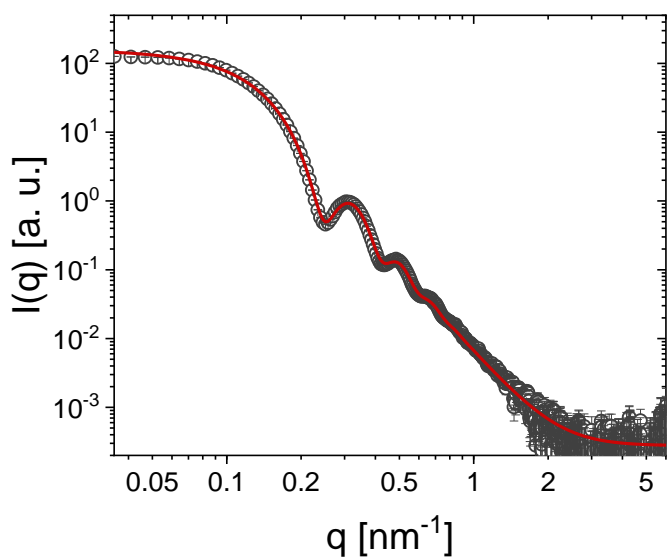

**Figure S3**. Scattering profile of the $SiO_2$ cores prior to the encapsulation in PNIPAM shells. The solid line is related to a hard sphere form factor fit.

**Transmission electron microscopy investigation**

The successful encapsulation of the silica cores and general morphology of the CS microgels was investigated with transmission electron microscopy (TEM). We used the images shown in **Figure S4a-c** to determine the encapsulation rate of the $SiO_2$ nanoparticles (NPs), the average core size and its size distribution which is presented in **Figure S4d**. The histogram is based on manual image analysis using the image analysis software imageJ.[7] The size distribution can be described with a Gaussian distribution function resulting in an average radius of 18 ± 2 nm.

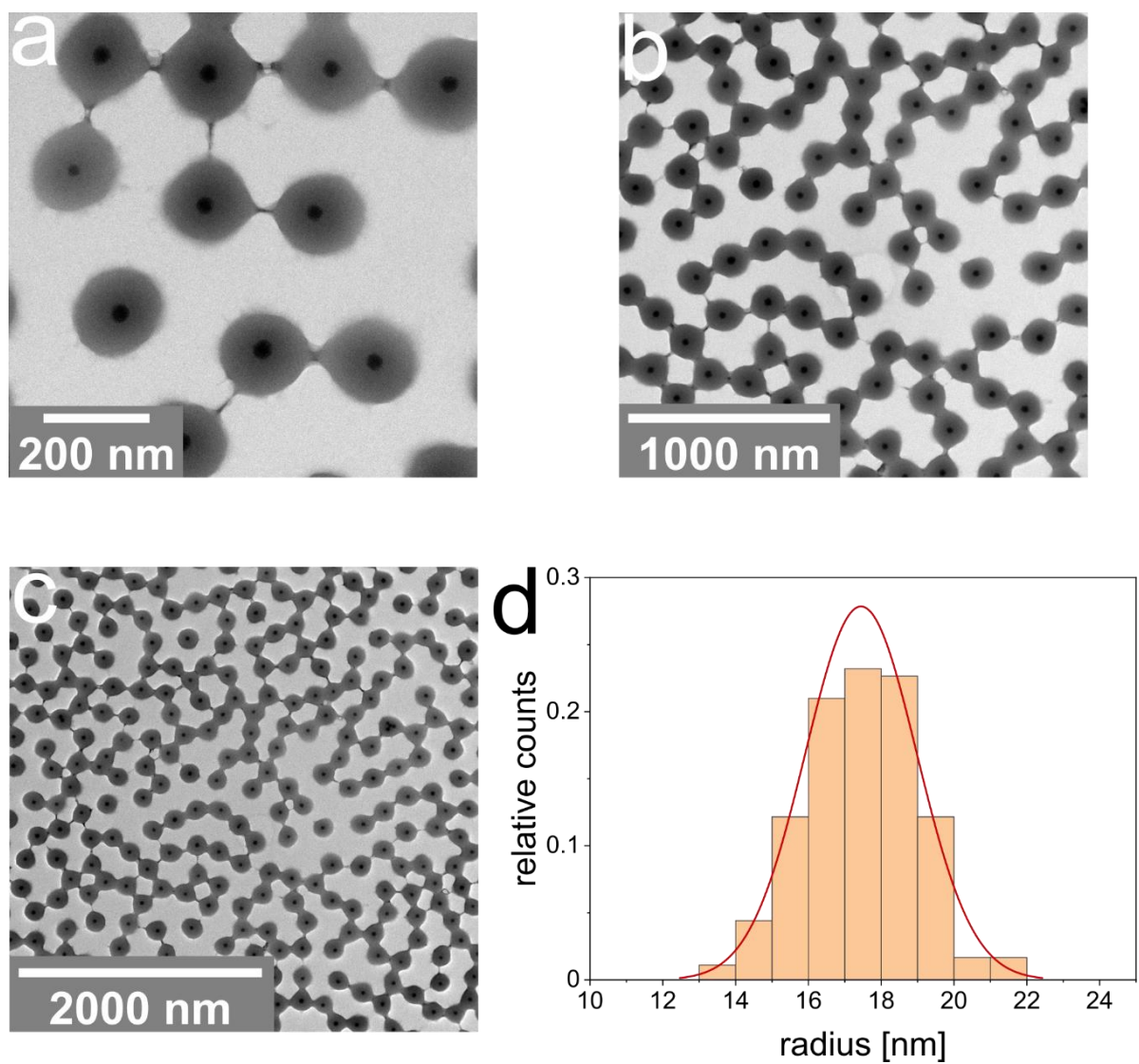

**Figure S4**. Representative TEM images of the CS microgels at different magnifications (a-c) and a histogram related to the size distribution of the $SiO_2$ cores with a fit to the data using a Gaussian distribution function (red line) (d).

**Sample annealing**

After sample preparation at room temperature, we performed a detailed annealing procedure where the temperature is increased from 20 °C to 50°C with a rate of 1.5 K/h using a high precision circulating water bath. At 50 °C the temperature is kept constant for one hour and is then again lowered to 20°C with a cooling rate of 1.5 K/h. The time dependent evolution of the temperature is shown in **Figure S5**.

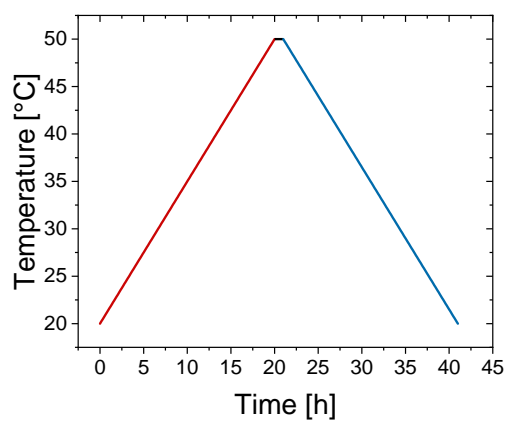

**Figure S5**. Temperature profile used for sample annealing.

**Vis-NIR absorbance spectra from different points in time**

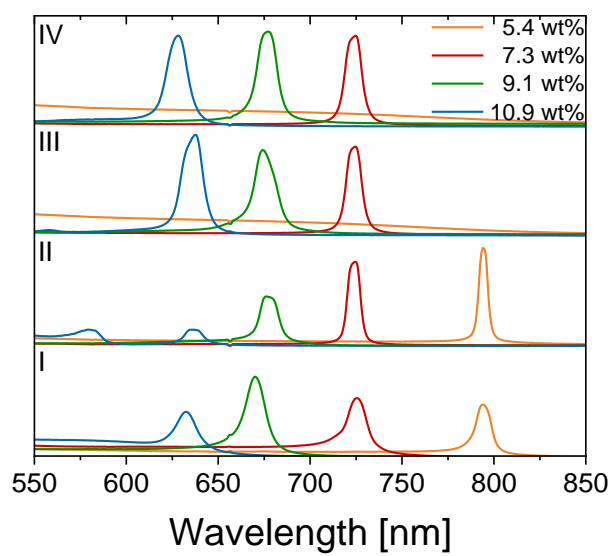

**Figure S6**. Vis-NIR absorbance spectra recorded at different points in time. From the bottom to the top: Directly after sample preparation (I), after annealing (II), half a year after sample preparation (III) and directly after SAXS measurements (IV).

Vis-NIR absorbance spectra were recorded at different points in time at 20°C, to verify the stability of the colloidal crystals (CC). The spectra at the bottom of **Figure S6** (I) were recorded directly after sample preparation and show Bragg peaks for all samples. After the annealing procedure (II) the spectra exhibit sharper Bragg peaks for the 5.4 and 7.3 wt% samples indicating an increased crystallinity. The samples with 9.1 and 10.9 wt% show a decrease in the intensity of the Bragg peak. We attribute this to the absence of large crystalline domains in the monitored sample volume. We note that the samples were measured in specific sample holders which do not allow for any changes in the position of the capillary. Afterwards samples were directly stored at 4°C. The next series of absorbance spectra (III) was recorded half a year after sample preparation and all samples excluding the 5.4 wt% exhibit distinct Bragg peaks. We conclude that the 5.4 wt% sample does not show stable colloidal crystals in long time equilibrium state. The improvement in the Bragg peak quality, in the observed volume, might be attributed to a temperature change of the samples during the transfer of the samples to a different location. Here, some rearrangements within the sample might occur resulting in higher quality Bragg peaks in the probed volume. Spectra recorded after the SAXS measurements (IV) show similar appearance compared to (III). A distinctive change in the Bragg peak position of the CS microgel dispersions is not detectable over the complete series of measurements.

**Crystal analysis by Vis-NIR and angle-dependent specular reflectance spectroscopy**

We measured concentrated dispersions of CS microgels that exhibit distinct and narrow Bragg peaks at wavelengths, $\lambda_{Bragg}$, that are related to the respective diffraction order $m$, the incident angle $\theta$ and the spacing between the respective lattice planes $d_{hkl}$. Theoretically, the Bragg peak position is described by a combination of Snell's law and Bragg's law:.[8]

$$m\lambda_{\text{diff}} = 2\, d_{\text{hkl}} \sqrt{n_{\text{crystal}}^2 - \sin\theta^2} \tag{S16}$$

We used an average refractive index for the CS microgel dispersion of $n_{\text{crystal}} = 1.345$ as reported in literature for a similar system.[9]

Exemplary angle-dependent specular reflectance spectroscopy measurements were conducted with the 9.1 wt% sample. In **Figure S7** we show the relation between the angle of the incident beam and the position of the Bragg peak. The black vertical lines indicate the theoretically expected Bragg peak positions for a $d_{hkl}$ = 251 nm. An absorbance spectrum recorded in transmission geometry to represent the Bragg peak position at an angle of 0° is shown in red. We see a distinct blue shift of the Bragg peak with an increase in the respective angle. The experimental data and the theoretical peak positions are in good agreement.

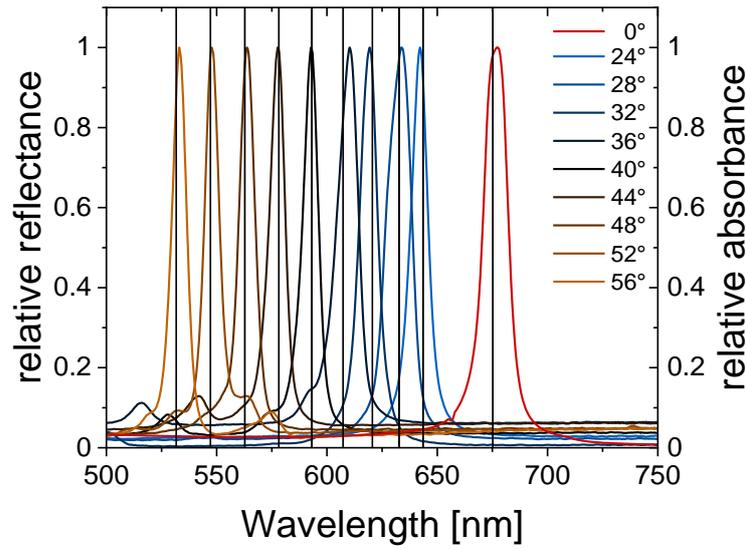

**Figure S7**. Normalized angle-dependent specular reflectance measured form the 9.1 wt% sample. The red spectrum is related to an absorbance measurement in transmission geometry (0 °). Black vertical lines indicate the theoretical peak positions based on a hcp lattice.

When the samples are measured in transmission geometry and $\theta$ corresponds to the angle between the optical normal and the incident beam $\theta$ = 0° leading to:

$$\lambda_{\text{Bragg}} = 2\, d_{\text{hkl}}\, n_{\text{crystal}} \tag{S17}$$

Assuming a hexagonal closed packed (hcp) crystal structure the spacing between lattice planes, $d_{hkl}$ is connected to the lattice parameters *a* and *c* by:

$$d_{hkl} = \frac{a}{\sqrt{\frac{4}{3}(h^2+hk+k^2)+\frac{a^2}{c^2}l^2}} \tag{S18}$$

Here, *h*, *k* and *l* correspond to the Miller indices. In case of a closed packed lattice of isotropic spheres the ratio between *c* and *a* is fixed to a value of $(8/3)^{1/2}$. This is due to the geometry of the unit cell at dense packing of spheres with a volume fraction of 0.74. With the known ratio we can rewrite **equation S18** to yield the lattice constant *a*:

$$a = d_{hkl} \sqrt{\frac{4}{3}(h^2 + hk + k^2) + \frac{3}{8}l^2} \tag{S19}$$

With the **equations S17** and **S19,** we can extract the lattice constant *a* from the position of the Bragg peak in the Vis-NIR absorbance spectrum.

$$a = \frac{\lambda_{\text{diff}} \sqrt{\frac{4}{3}(h^2 + hk + k^2) + \frac{3}{8}l^2}}{2\, n_{\text{crystal}}} \tag{S20}$$

When we assign the 002 plane to the Bragg peak, we can write **equation S20** in the following way:

$$a = \frac{\lambda_{\text{diff}} \sqrt{\frac{3}{2}}}{2\, n_{\text{crystal}}} \tag{S21}$$

The volume fraction $\phi$ of an hcp crystal with 3 + 3 spheres contributing to one unit cell can be calculated with the lattice constant *a* and the radius *R* of the spheres.

$$\phi = \frac{(3+3)\frac{4}{3}\pi R^3}{\frac{3\sqrt{3}\sqrt{\frac{8}{3}}a^3}{2}} \tag{S22}$$

**Temperature dependent Vis-NIR absorbance spectroscopy**

We performed temperature dependent Vis-NIR spectroscopy with crystalline CS microgel dispersions which exhibit Bragg peaks. The spectra were recorded between 20 °C and 50 °C in steps of 0.3 °C. The samples were equilibrated for twelve minutes before each measurement. Due to the slow heating and cooling rates, we can assume that the sample is close to equilibrium conditions during the whole procedure. This is illustrated in **Figure S8a**, while in **Figure S8b** a small section of the procedure is shown to clarify the stepwise approach. We note that **FigureS8** excludes the time needed for the sample holder to change temperature, and the time for conducting the measurement itself.

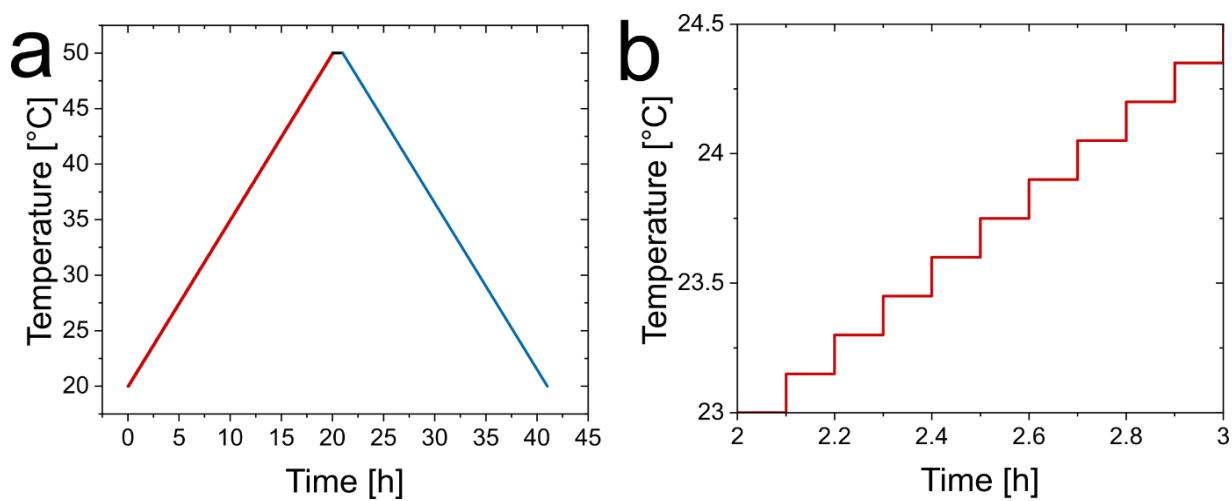

**Figure S8**. Temperature profiles used for Vis-NIR absorbance measurements. (a) Time dependent evolution of the temperature and a small section (b) showing the stepwise increase of the temperature.

The temperature dependent Vis-NIR spectra are shown in **Figure S9** and were recorded between 20 °C and 50 °C in 0.3 °C steps indicated by the color transition from dark blue (20 °C) to red (50 °C). The four rows correspond to the mass contents of 7.3 wt% (a), 9.1 wt% (b, c) and 10.9 wt% (d) CS microgel dispersions. Noteworthy, for 9.1 wt% an additional sample was measured to verify the reproducibility of the measurements. For each sample, we observe a decrease in the absorbance of the Bragg peak in a certain temperature range, which is visible in the absorbance spectra and marked in green for the respective plots of the Bragg absorbance as function of the temperature. In this temperature range, the sample undergoes the phase transition from a crystalline to a disordered system. Based on the conducted experiment, we can extract the transition temperature of the respective CS microgel dispersion (**Table S4**). This is done by applying linear fits on the green colored transition section were the Absorbance of the Bragg peak decreases and the blue colored section where the Bragg peak absorbance is 0. The intercept of both fits on the x-axis is listed as transition temperature.

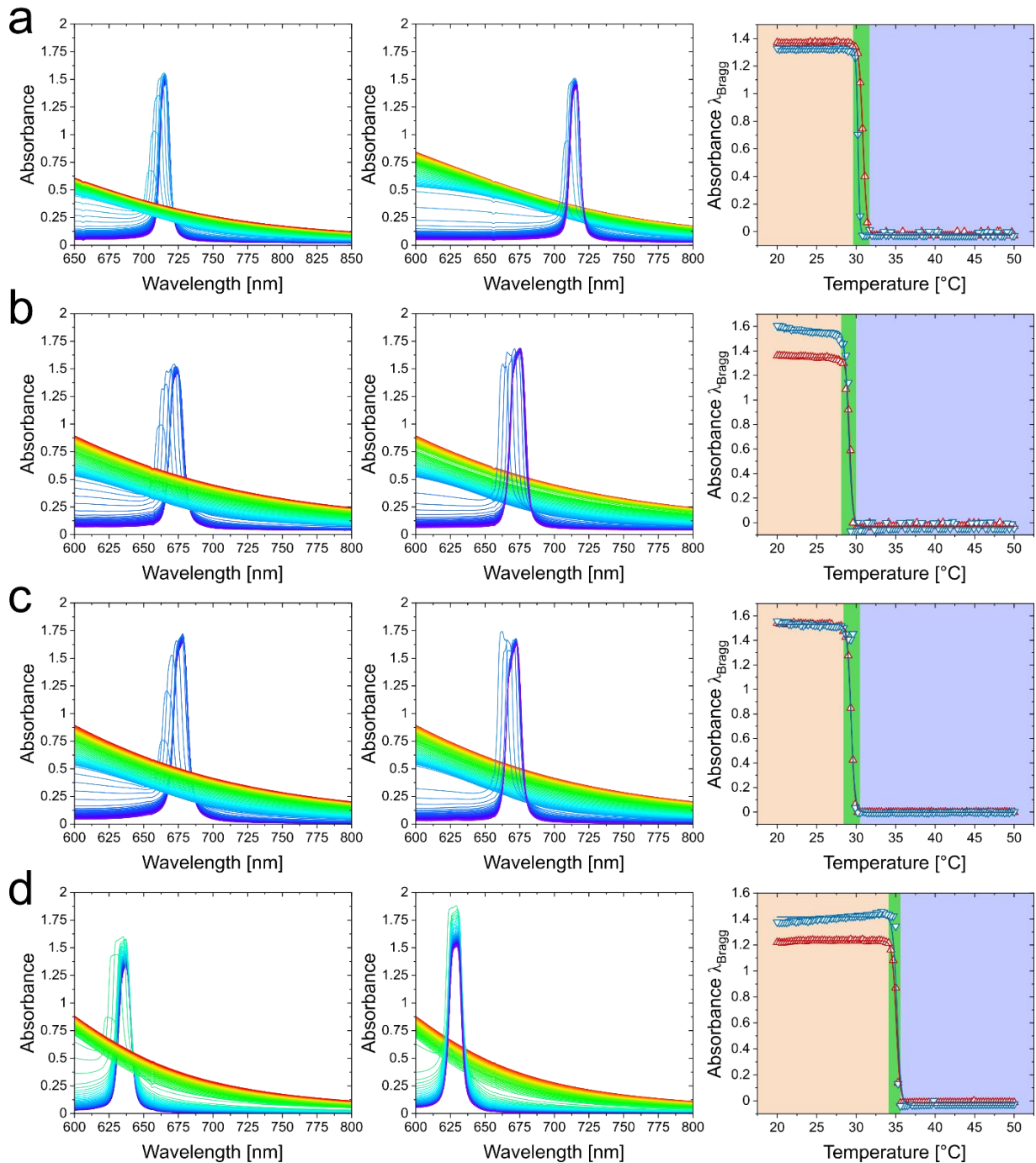

**Figure S9.** Temperature dependent Vis-NIR absorbance spectra and the respective Bragg peak intensity as function of the temperature from 7.3 wt% (a), 9.1 wt% (b, c) and 10.9 wt% CS microgel dispersion. The spectra in the left row correspond to a heating cycle while the spectra in the middle column correspond to a cooling cycle. Spectra were recorded from 20°C to 50°C and vice versa with 0.3°C steps indicated by the color transition from blue to red.

**Table S4**. Transition temperatures of the respective CS samples.

| Mass content [wt%] | T$_{transition}$ (heating) [°C] | T$_{transition}$ (cooling) [°C] |
|---|---|---|
| 7.3 | 31.6 | 30.7 |
| 9.1 | 29.8 | 29.9 |
| 9.1 | 30.2 | 30.3 |
| 10.9 | 35.7 | 35.5 |

**Density of the silica cores**

In order to extract the number concentration from SAXS profiles the density of the scattering objects is required. In our case, the form factor of the core is used for the analysis and thus the density of the silica cores was determined. To do so, the dispersion of silica cores was purified via dialysis against ultra-pure water and the density of a series of dilute dispersions with various concentrations was measured. From a plot of the reciprocal density of the respective dispersions against the respective mass content, the density of the SiO$_2$ NPs can be extracted with the following equation:

$$\rho = \frac{1}{m+b} \implies \frac{1}{\rho} = m + b \tag{S23}$$

Here, $\rho$ is the density of the dispersed particles and $m$ and $b$ are related to the slope and the y-intercept of the linear fit as shown in **Figure S10**.

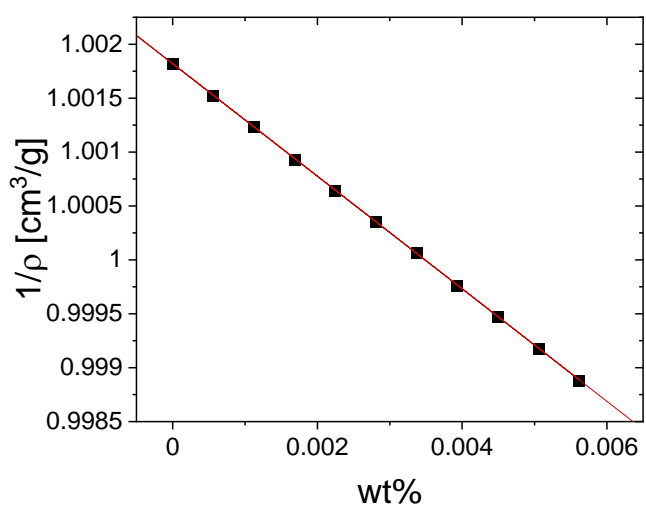

**Figure S10**. Reciprocal density of dilute SiO$_2$ NP dispersions against the mass content of the respective dispersion. The red line corresponds to the linear function used to fit the data.

The linear function describes the data very well and a slope of -0.52204 cm$^3$/g and a y-intercept of 1.00182 cm$^3$/g is obtained. According to equation S15 a density of 2.08 g/cm$^3$ for the SiO$_2$ NPs can be determined.

**Extraction of number concentrations from SAXS**

Additional SAXS measurements were performed with the intensity in absolute units on four (I-IV) concentrated CS microgel dispersions to determine the number concentrations based on the scattering contribution of the SiO$_2$ cores.[10, 11] A total number of four samples with the same mass content were investigated to ensure that the performed measurements lead to representative results and to obtain a standard deviation regarding the number concentration. The SAXS profiles of the CS microgel dispersions (10 wt%, H$_2$O), including form factor fits (solid lines) are shown in **Figure S11**. The applied polydisperse sphere model describes the data sufficiently well in the respective $q$-range and the fit parameters are listed in **Table S5**. We see the first form factor minimum at approximately 0.2 nm$^{-1}$ and discrepancies between the measured data and the form factor fit below $q$ = 0.06 nm$^{-1}$. The deviation is related to scattering of the shell which is not described by the model used here. The data of the four different CS microgel dispersions superimpose each other, as well as the form factor fits. This is expected as all samples possess the same concentration. Even though 10 wt% is a concentration high enough that strong structure factor contributions will occur, our analysis is reasonable since we focus on the form factor of the cores that are much smaller than the total CS microgel dimensions. In other words, the structure factor from the lattice of CS microgels and the form factor of the small cores appear in different regimes of $q$. In particular the pronounced form factor oscillations visible in **Figure S11** are not affected by the structure factor. Therefore, we can use the analysis of relatively high and thus well strongly scattering samples to deduce the number concentration of cores and consequently the CS microgels based on the assumption that each CS microgel contains a single silica core. Consequently, the number concentration of the cores is the same as the number concentration of the CS microgels.

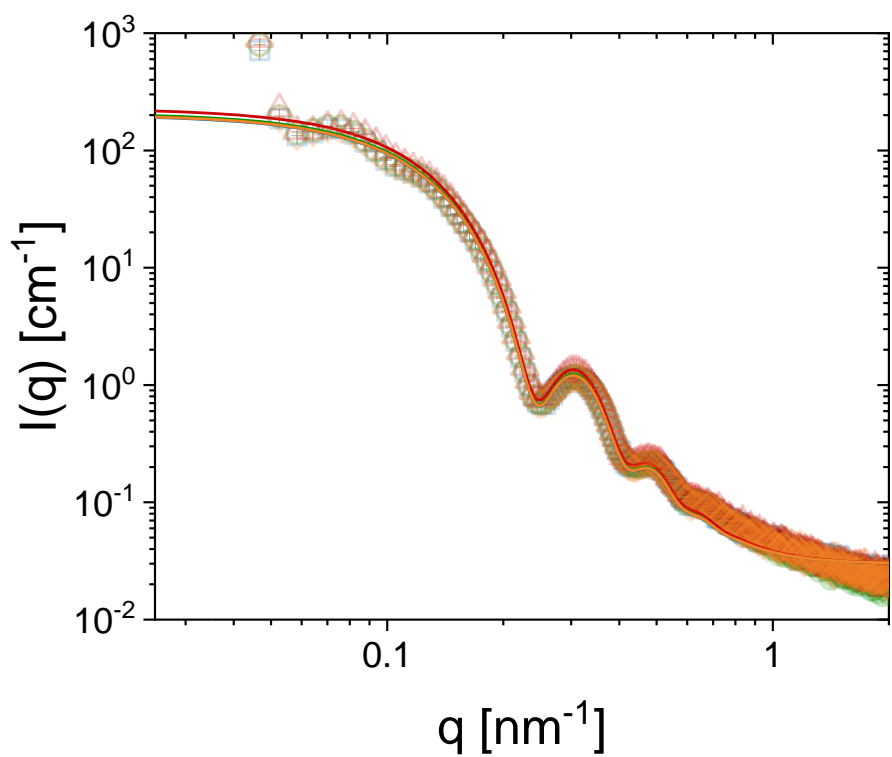

**Figure S11.** SAXS profiles of four different samples (each 10 wt% CS microgels) and the respective form factor fits to the core scattering contribution (solid lines).

**Table S5.** Fit parameters obtained from the form factor fits. The Errors of the parameters are smaller by two orders of magnitudes compared to the obtained valuers and therefore not listed.

|  | I | II | III | IV |
|---|---|---|---|---|
| *scale* | 0.00107 | 0.00108 | 0.00111 | 0.00121 |
| $I_B$ [cm$^{-1}$] | 0.03 | 0.03 | 0.03 | 0.03 |
| SLD (SiO$_2$) [10$^{-6}$ Å$^2$] | 17.75 | 17.75 | 17.75 | 17.75 |
| SLD (H$_2$O) [10$^{-6}$ Å$^2$] | 9.47 | 9.47 | 9.47 | 9.47 |
| R [nm] | 18 | 18 | 18 | 18 |
| $\sigma_{poly}$ | 0.1 | 0.1 | 0.1 | 0.1 |

In order to calculate the number concentration *N*, the scattering intensity at infinitely small *q*, $I_0$ needs to be known. Therefore, the SAXS data in absolute units are plotted in a Guinier plot, *i.e.* ln(*I*(*q*)) as function of $q^2$ according to:

$$\ln(I(q)) = \ln(I_0) - \frac{q^2 R_G^2}{3} \tag{S24}$$

In addition to $I_0$, we can extract the radius of gyration $R_G$ from a linear fit to the data. As presented in **Figure S12**, we performed the Guinier analysis not only on the scattering profile but also on the performed form factor fits of the core. This is due to the form factor oscillations of the shell, which occur for $q^2 < 0.015$ nm$^{-2}$ and might influence the results of the Guinier analysis. To prevent possible corruption of the extracted $I_0$ due to the scattering of the shell we used the $I_0$ extracted from the Guinier analysis of the form factor fits of the core. The results from the Guinier analysis are presented in **Table S6** and **Table S7**.

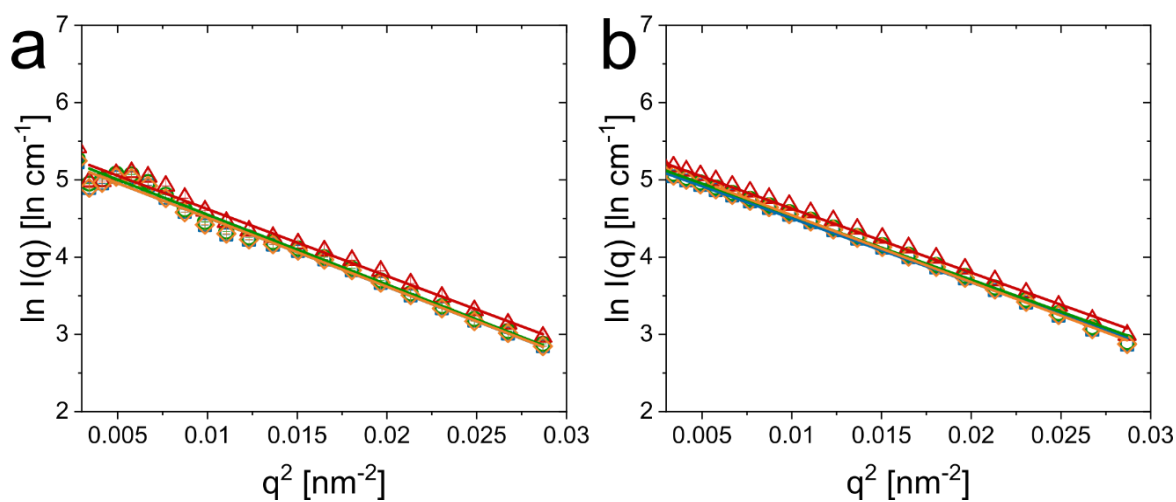

**Figure S12**. Guinier plots of the scattering data (a) and the respective form factor fits (b).

**Table S6**. Results from the linear fits applied on the Guinier plots of the SAXS data.

|  | I | II | III | IV |
|---|---|---|---|---|
| $I_0$ [cm$^{-1}$] | 221 ± 10 | 233 ± 10 | 242 ± 10 | 221 ± 9 |
| $R_G$ [nm] | 16.4 ± 0.3 | 16.5 ± 0.3 | 16.1 ± 0.3 | 4 ± 0.3 |

**Table S7**. Results from the linear fits applied on the Guinier plots of the form factor fits.

|  | I | II | III | IV |
|---|---|---|---|---|
| $I_0$ [cm$^{-1}$] | 206 ± 2 | 215 ± 2 | 235 ± 2 | 224 ± 3 |
| $R_G$ [nm] | 15.8 ± 0.1 | 15.8 ± 0.1 | 15.8 ± 0.1 | 16.2 ± 0.1 |

Both approaches give comparable results with a radius of gyration close to 16 nm. The radius of gyration is expected to follow $R_g = \sqrt{\frac{3}{5}} R_{HS}$ and, therefore, to be smaller than the radius we extracted from form factor fits. With a $R_g$ of 16 nm we see a small deviation of 2 nm from the expected value. This is attributed to the selected q-range where the Guinier analysis was performed due to limitation in the experimentally accessible lowest *q*. Regarding $I_0$, we obtained slightly higher intensities from the linear fits applied on the Guinier plot of the experimental data. This is due to the scattering intensity in the low *q*-regime is not only related to the scattering of the core. Here, the form factor oscillations of the shell are interfering with the scattering signal of the core in the selected *q*-regime. Thus, the measured intensities are influenced by the scattering of the shell and therefore lead to higher scattering intensities at infinitesimal low *q*.

Since we now know the forward scattering intensity, $I_0$, we can combine this with the Avogadro's number $N_A$, the density of the silica cores *ρ*, the mass of a single scattering object, *m*, the molecular weight, of the scattering object, $M_w$, and the difference in scattering length density between the solvent and the scattering object ΔSLD to calculate the number concentration *N* according to **equation S25**:

$$N = \frac{I_0 \, N_A \, \rho^2}{m \, M_w \, \Delta SLD^2} \tag{S25}$$

The extracted number concentrations, as well as the average number concentration $\bar{\bar{N}}$ and the respective standard deviation *SD* are listed in **table S8**.

**Table S8**. Number concentrations extracted from the scattering intensity of CS microgels dispersions (10wt%)

|  | I | II | III | IV | $\bar{N}$ | SD |
|---|---|---|---|---|---|---|
| N (*I(q)*) [$10^{13}$ 1/mL] | 5.31 | 5.58 | 5.81 | 5.31 | 5.50 | 0.21 |
| N (*P(q)*) [$10^{13}$ 1/mL] | 4.95 | 5.15 | 5.64 | 5.36 | 5.27 | 0.26 |

The average number concentration is 5.5 ± 0.2 $10^{12}$ 1/mL and is, within the experimental errors, virtually the same as the number concentration extracted from the form factor fits with 5.3 ± 0.3 $10^{12}$ 1/mL. We ascribe the difference between the two values due to the contributions of the polymer shell on the *P(q)* and, therefore, we will use the number concentration extracted from the form factor fits for further calculations.

From the number concentration and the hydrodynamic radius $R_h$ of the CS microgels we can calculate the generalized volume fraction ξ:

$$\xi = N\frac{4}{3}\pi R_h^{\,3}(T) \tag{S26}$$

Due to the PNIPAM shell, the hydrodynamic radius and the generalized volume fraction of the CS microgels depend on the temperature as shown in **Figure S13**.

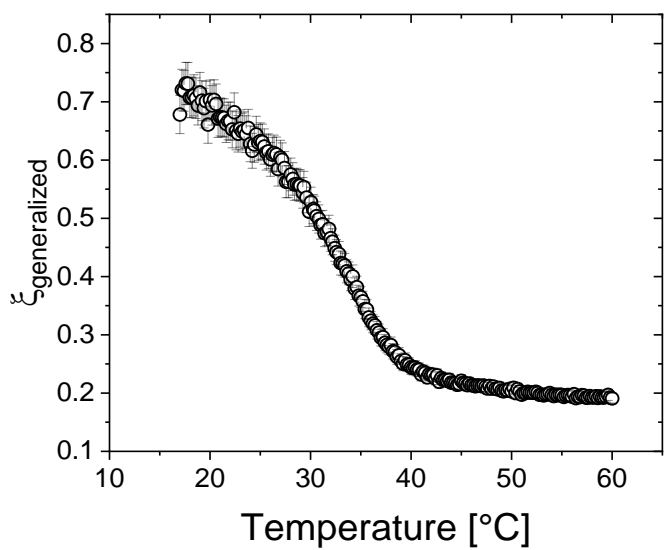

**Figure S13**. Generalized volume fraction as function of the temperature on the example of a 10 wt% CS microgels dispersion.

The hydrodynamic radius, measured via temperature dependent DLS, directly corresponds to the relation of the generalized volume fraction and the temperature.

**Structure factor extraction in the fluid regime**

Absorbance spectroscopy has revealed that all CS microgel dispersions (5.4 – 12 wt%) are not in a crystalline state at 40 °C. Thus, we assume that the sample possess a fluid-like structure. To verify this, we conducted SAXS measurements at 40 °C. The 2D detector images of all concentrated samples are shown in **Figure S14**. All scattering patterns exhibiting a fluid-like structure factor contribution as well as intensity minima related to the form factor of the collapsed CS microgels. No Bragg peaks were observed so the dispersions are in a purely unordered phase.

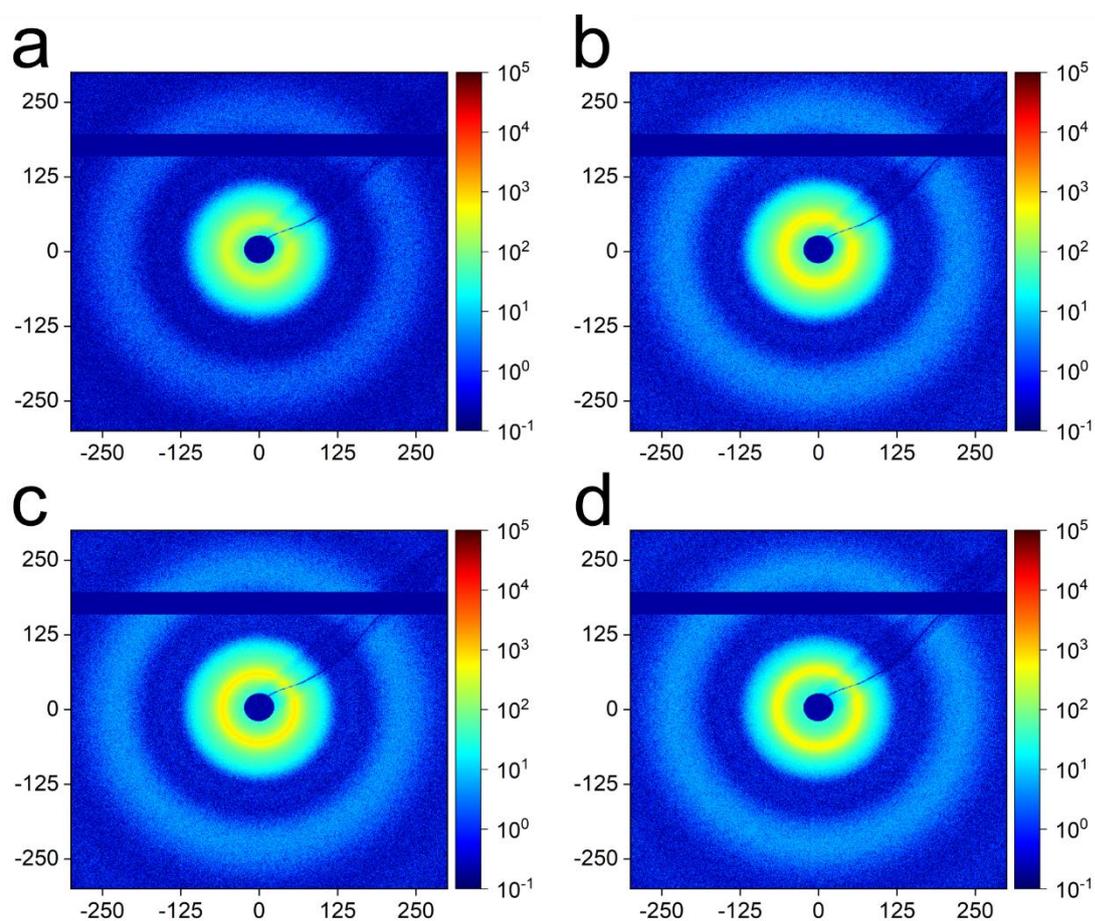

**Figure S14.** 2D detector images of CS microgel dispersions with mass contents of 5.4 to 12 wt%.

Radially averaged scattering profiles of the CS microgels with mass contents between 5.4 and 12 wt% recorded at 40 °C are shown in **Figure S15**. While the form factor at mid to high $q$ remains unchanged for increasing concentration, the first structure factor maximum in the low $q$ region shifts towards higher $q$ as indicated by the grey arrow.

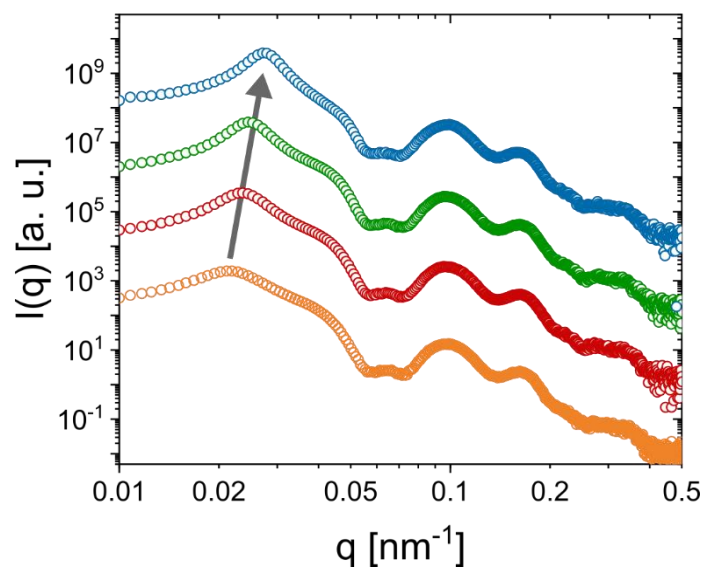

**Figure S15.** Scattering profiles of samples with 5.4 wt% (orange circles), 7.3 wt% (red circles), 9.1 wt% (green circles) and 12 wt% (blue circles, $H_2O$) recorded at 40 °C. The grey arrow indicates the shift of the structure factor maximum towards higher $q$ with an increase in mass content.

We also measured a sample at a much lower concentration (1.25 wt%) at 40 °C, i.e. at conditions where the microgel shell is collapsed. To our surprise the scattering pattern shown in **Figure S16** exhibits a weak structure factor maximum at $q ≈ 0.02$ nm despite the low concentration. Due to this, we did not use the complete scattering profile for the extraction of the structure factor from scattering profiles recorded of the dense samples. The structure factor contribution can be described by the Percus-Yevick hard sphere structure factor model resulting in a volume fraction of 0.16 and a hard sphere radius of 176 nm.

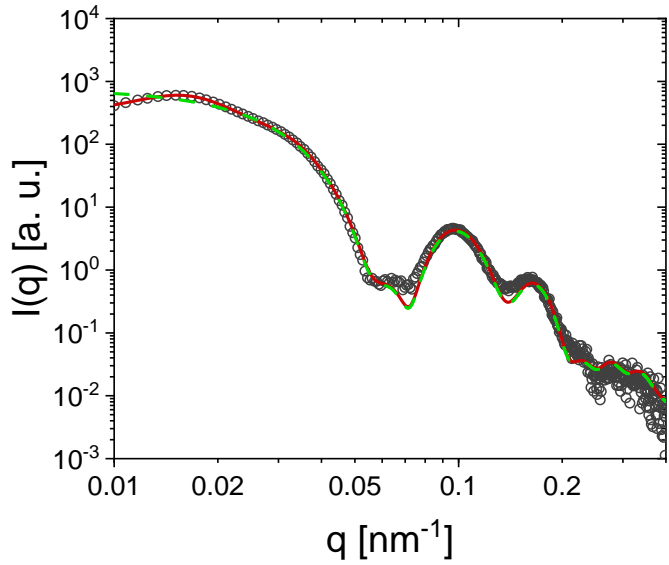

**Figure S16**. Scattering profile (black circles) recorded from a 1.25 wt% CS microgel dispersion at 40°C. The solid red line corresponds to a combination of form and structure factor fit including a core-homogeneous-shell model and the Percus-Yevick model, respectively. The green dashed line represents a form factor fit without any structure factor.

The structure factor can be calculated from the scattering profiles of the concentrated samples $I_{conc}(q)$ using the known form factor $P(q)$:

$$S(q) = \frac{I_{conc}(q)}{P(q)} \tag{S27}$$

Due to small deviations of the form factor fit and the oscillations of the scattering profile, the extracted structure factors show distinct deviation from unity in the mid to high range of $q$. Here, the deviations are present even for $q > 0.05$ nm$^{-1}$ where we expect only low amplitudes of the oscillations related to the structure factor. In order to conduct a more realistic structure factor extraction, we used the form factor fit for the extraction in the low q regime ($q < 0.025$ nm$^{-1}$) which is shown as green dashed line in **Figure S16** and the scattering profile of the dilute CS microgel dispersion for the mid and high q-regime ($q > 0.025$ nm$^{-1}$) as indicated in **Figure S17b**.

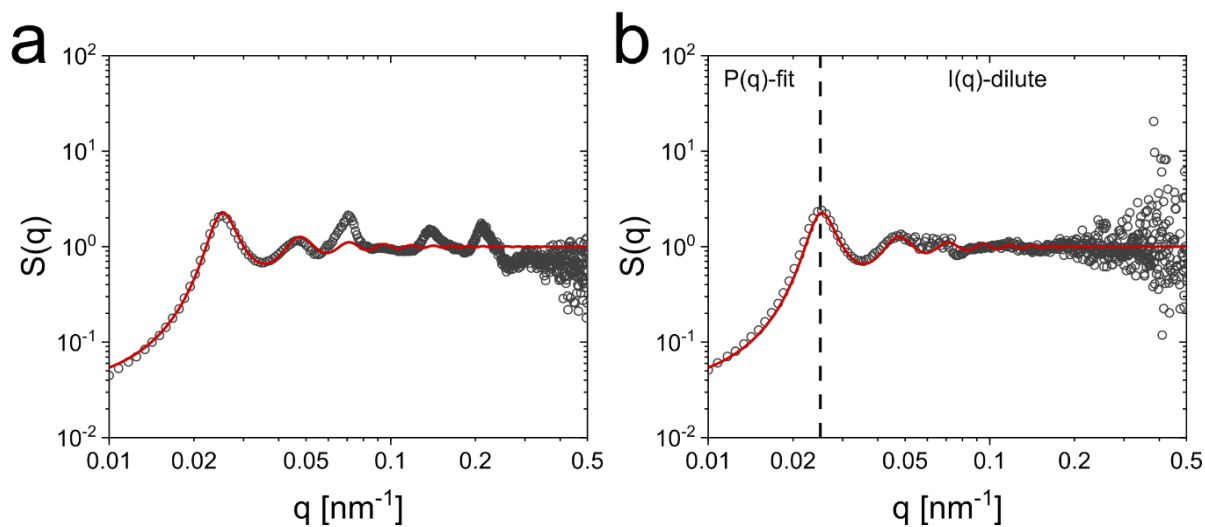

**Figure S17**. Extracted structure factors from a 9.1 wt% CS microgel dispersion recorded at 40 °C. (a) Extraction based on the form factor fit obtained from fitting the scattering profile of the dilute CS microgel dispersion. (b) Extraction based on the form factor fit for $q < 0.025$ nm$^{-1}$ and the scattering profile of the CS microgel in the dilute state for $q > 0.025$ nm$^{-1}$. The red lines show the structure factor fit (Percus-Yevick).

We fitted the extracted structure factors with the Percus-Yevick hard sphere model. The resulting volume fractions and hard sphere radii are listed in **Table S9**.

**Table S9**. Parameters obtained from structure factor fits applied to the scattering profiles recorded from CS microgel dispersions at 40°C.

|  | **5.4 wt%** | **7.3 wt%** | **9.1 wt%** | **12 wt%** |
| --- | --- | --- | --- | --- |
| $R_{HS}$ [nm] | 142 ± 11 | 135 ± 6 | 133 ± 5 | 124 ± 4 |
| $\phi_{HS}$ | 0.32 ± 0.08 | 0.38 ± 0.06 | 0.43 ± 0.05 | 0.46 ± 0.04 |
| $S(q)_{q=0}$ | 0.08 | 0.05 | 0.03 | 0.023 |

**Figure S18** shows the extracted structure factors along with the PY structure factor fits in the regime of low $q$ where deviations between fit and data are expected for soft particles. In our case, fits and data match also in the low $q$ regime indicating that our collapsed CS microgels effectively interact like hard spheres.

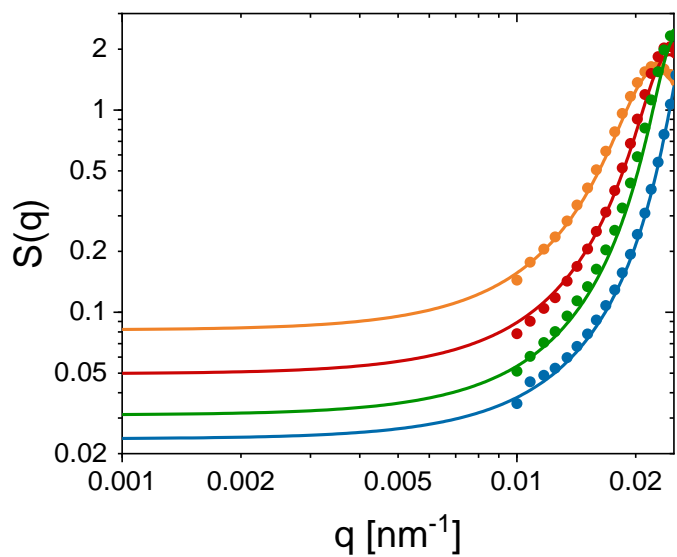

**Figure S18.** Low *q* region of experimentally determined structure factors (symbols) and Percus-Yevick fits (solid lines) from SAXS measurements of 5.4 wt% (orange dots), 7.3 wt% (red dots), 9.1 wt% (green dots) and 12 wt% (blue dots) samples at 40 °C.

**SAXS patterns and simulations for CS microgels in the solid (crystalline) state**

The recorded 2D detector images of samples with 5.4, 7.3, 9.1 and 10.9 wt% measured at 20 °C (swollen state) are presented in **Figure S19**. For the 6 wt% sample, at the first glance, we see an isotropic scattering pattern in **Figure S19a**, indicating a fluid-like structure. At a closer look, diffraction peaks are visible close to the beamstop, i.e. at low *q*. The large number of peaks and their rather random azimuthal distribution points to a multicrystalline character of the sample. This indicates that this sample is in the fluid-crystalline coexistence region with small, randomly oriented crystallites. Due to the dominant presence of a fluid-like structure factor contribution we consider the 5.4 wt% sample primarily as a fluid, which is in agreement to our findings from absorbance spectroscopy (see main manuscript). For the 7.3, 9.1 and 10.9 wt% samples, we see pronounced sharp Bragg peaks of multiple diffraction orders in **Figure S19b-d**. The hexagonal symmetry of these diffraction patterns indicates close packed structures. The large number of diffraction orders indicates long range order and thus large crystalline domains.

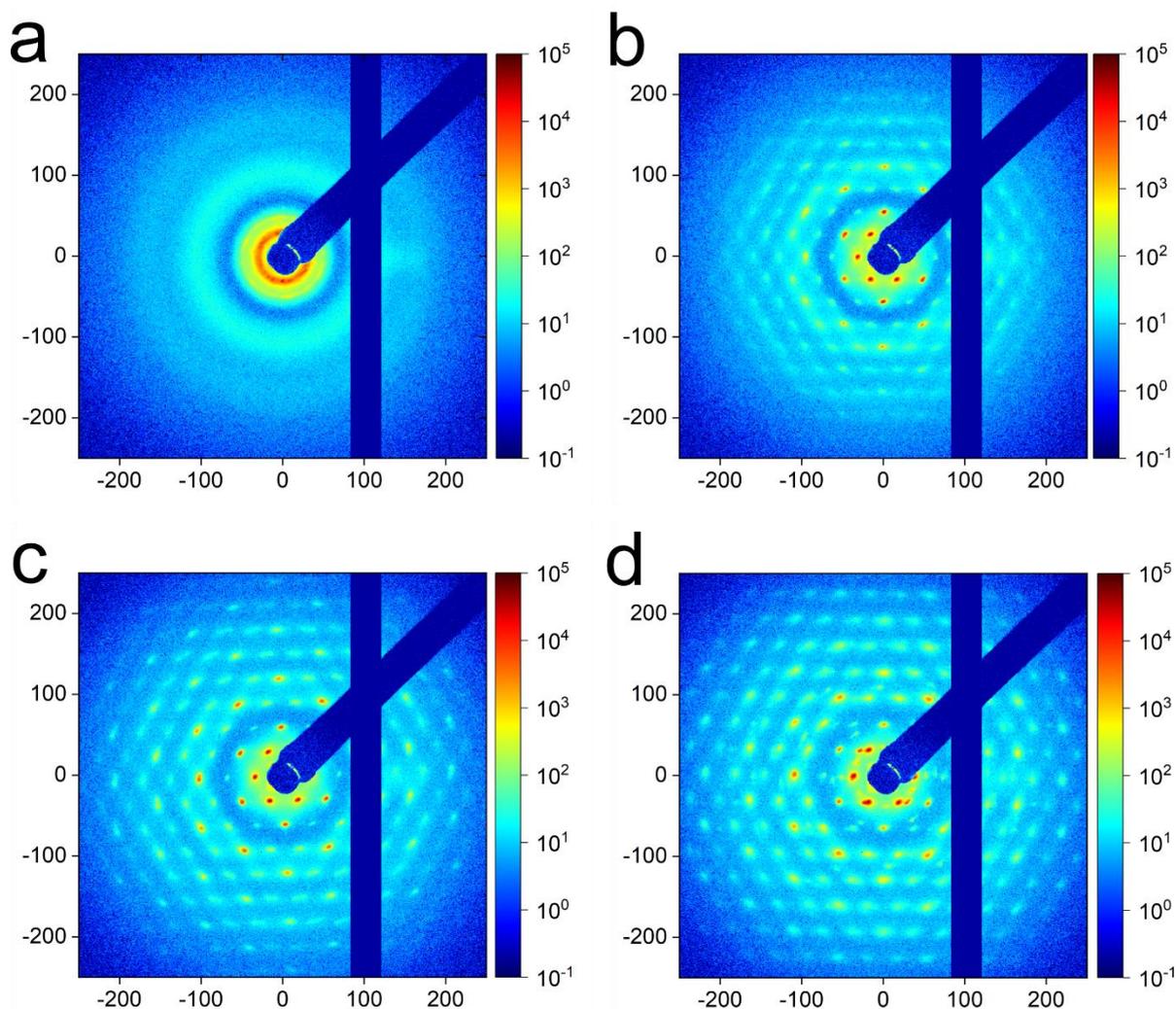

**Figure S19**. 2D detector images of samples with 5.4 wt% (a), 7.3 wt% (b), 9.1 wt% (c) and 10.9 wt% (d) CS microgels.

In order to identify the crystal structure, we simulated scattering patterns using the software *scatter*.[12] We see an excellent agreement between the recorded (left half) and the simulated pattern (right half) for an hcp structure in **Figure S20**.

The simulations were performed with the 002 plane of the hcp crystal positioned orthogonal to the incident beam. The parameters used in simulations are listed in **Table S10**. We took into account a small particle displacement of 5 nm within the lattices. In **Table S10**, "max. hkl" refers to the orders of Bragg peaks attempted to simulate and $\rho$ refers to the ratio in scattering length density between the core and the polymer shell. For simplicity, we used a core-shell model with a homogeneous shell to describe the form factor of the CS microgels. Differences in $R_{SAXS}$ used in this simulation and $R_{SAXS}$ obtained from the form factor analysis are ascribed to the different models used to describe the form factor of the CS microgels.

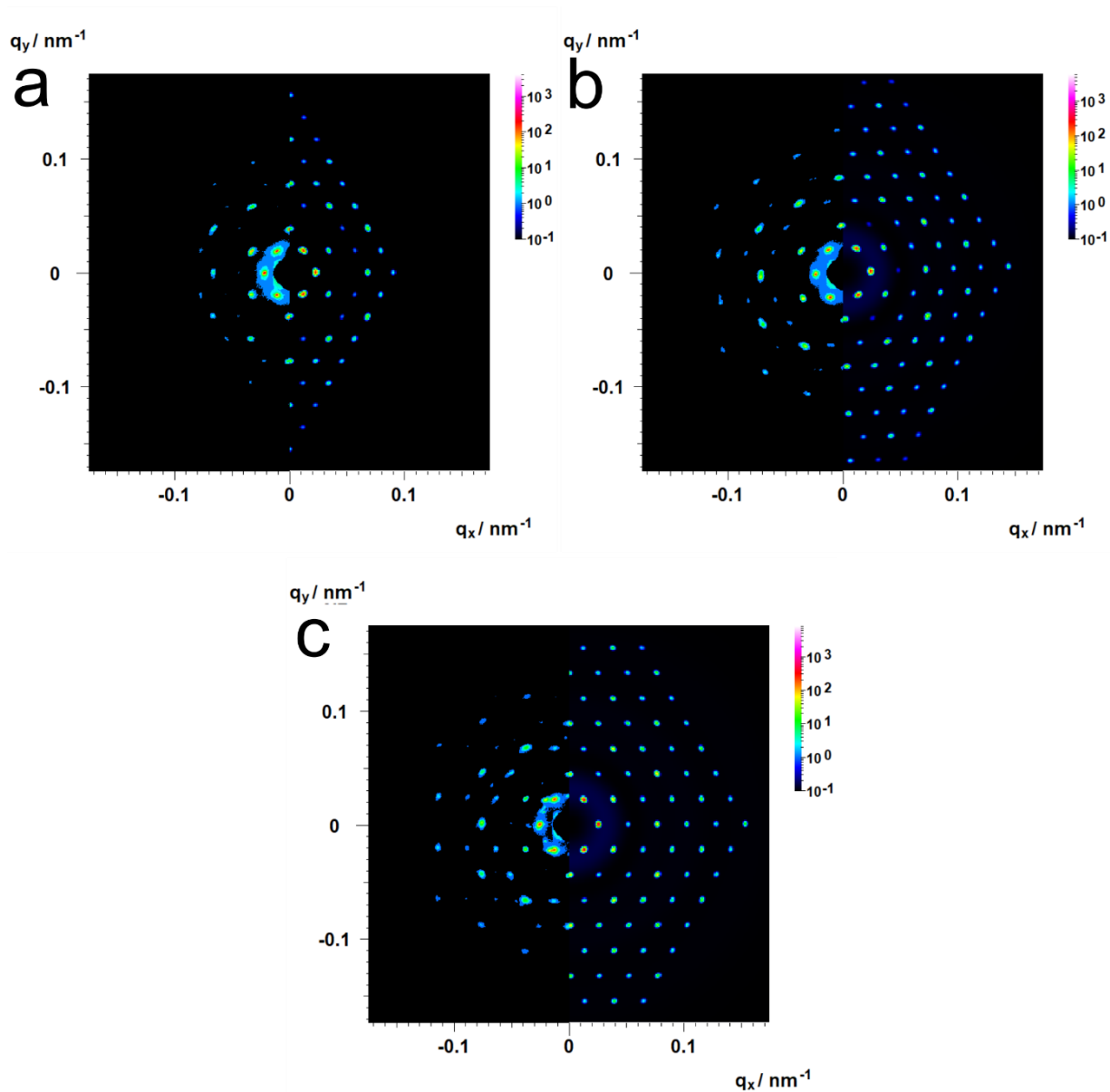

**Figure S20**. Experimental (left half) and simulated (right half) scattering patterns of samples with 7.3 wt% (a), 9.1 wt% (b) and 10.9 wt% (c) CS microgels. Simulations were performed using a hcp crystal structure with the beam direction orthogonal to the 002 plane.

**Table S10**. Parameters used to simulate the scattering pattern with a hcp crystal structure.

|  | 7.3 wt% | 9.1 wt% | 10.9 wt% |
| --- | --- | --- | --- |
| unit cell [nm] | 322 | 303 | 284 |
| radial domain size [nm] | 2700 | 3000 | 3200 |
| azimuthal domain size [nm] | 2000 | 2200 | 2500 |
| displacement [nm] | 5 | 5 | 5 |
| max. hkl | 4 | 6 | 6 |
| $R_{core}$ [nm] | 18 | 18 | 18 |
| $\sigma_{core}$ | 0.1 | 0.1 | 0.1 |
| $R_{SAXS}$ [nm] | 121 | 114 | 104 |
| $\rho$ | 0.04 | 0.04 | 0.04 |

**Comparison of peak positions for close packed crystal structures**

We now want to compare our detected Bragg peaks with the theoretically allowed ones including other crystal structures than hcp. For this, we use the extracted structure factors from the 9.1 wt% sample as an example. The structure factors were extracted by dividing the measured scattering intensity of the dense sample by the form factor contribution of the silica cores (green circles) as well as the form factor of the total CS microgel (blue circles). We plotted all possible peaks corresponding to a hcp structure as vertical lines in the normalized scattering profiles in **Figure S21a**. The profiles were normalized to the position of the first structure factor maximum. The red lines correspond to the Bragg peaks that we could detect in our measured profiles while the black lines indicate the position of additional, theoretically allowed, peaks. The peak position is not influenced by the protocol used for the extraction of the structure factor. This is clearly shown as the peak maxima of all three datasets are located on the same positions in $q$.

Since the fcc and hcp crystal structure are energetically very similar and both structures were reported for dense phases of soft microgels, it is worth to compare these to the measured scattering profiles. In **Figure S21b**, the redlines indicate the theoretical position of peaks related to an hcp, the black lines for fcc, and the green lines for peaks shared by both crystal structures. The black lines do not match with the positions of the recorded Bragg peaks and even for peak positions shared between fcc and hcp structures (green) we only see a match between experimental peaks and theoretical positions at $q/q^* = 1.73$ and 3. For a rhcp structure, we would expect to detect at least some Bragg peaks solely attributed to an fcc structure. Thus, we can exclude a fcc structure and find no indications for the presence of a rhcp structure. Here we want to note that the $q/q^*$ normalization regarding the fcc peak positions needed to be performed on the Bragg peak related to the (220) plane. This peak is present in both fcc and hcp where it corresponds to the (110) plane. Therefore, we selected the peak at $q/q^* = 1.73$.

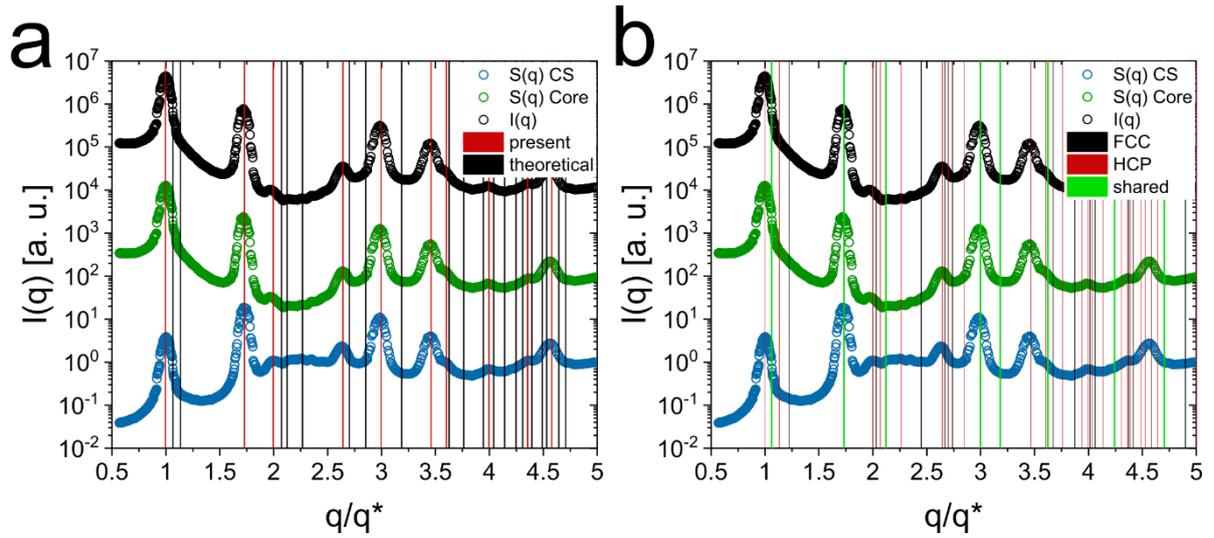

**Figure S21**. Scattering profiles (black circles) and the extracted structure factors using the form factor of just the core (green circles) and of the total CS microgel (blue circles). (a) The vertical lines correspond to the theoretically allowed peak positions for an hcp crystal structure. Red lines are related to the experimentally observed peaks while the black lines indicate the position of additional theoretically allowed peaks. (b) Red lines correspond to peak positions of an hcp structure, black lines are related to an fcc structure and the green lines indicate peaks shared in position between both close packed structures.

**CS microgel dispersion close to the freezing volume fraction**

The 5.4 wt% CS microgel sample exhibits a distinct fluid-like structure factor, but also a Bragg peak around $q \approx 0.03$ nm$^{-1}$ indicating the coexistence with crystallites. In addition to the modeled form factor, we applied a structure factor fit to describe the fluid-like contribution (Percus-Yevick, $S_{\text{fluid}}(q)$).[13] Here, we want to note that we applied a decoupling approximation instead of a local monodisperse approximation regarding the calculation of the structure factor.[14] Thus, we take into account the polydispersity of our system. The analysis yields a hard sphere radius of 158 nm and a volume fraction $\phi$ of 0.44. In **Figure S22** we see a good agreement between the fit (black line) and the scattering profile (orange circles). We attribute the deviation between $R_{\text{SAXS}}$ and the hard sphere radius to the fuzziness of the microgel shell and potential electrostatic interaction between the CS microgels. The structure factor related to the presence of small crystalline residuals ($S_{\text{crystal}}(q)$) was extracted according to:

$$S_{\text{crystal}}(q) = \frac{I(q)}{N(\Delta SLD)^2 V_p^2 P(q) S(q)_{\text{fluid}}} \tag{S28}$$

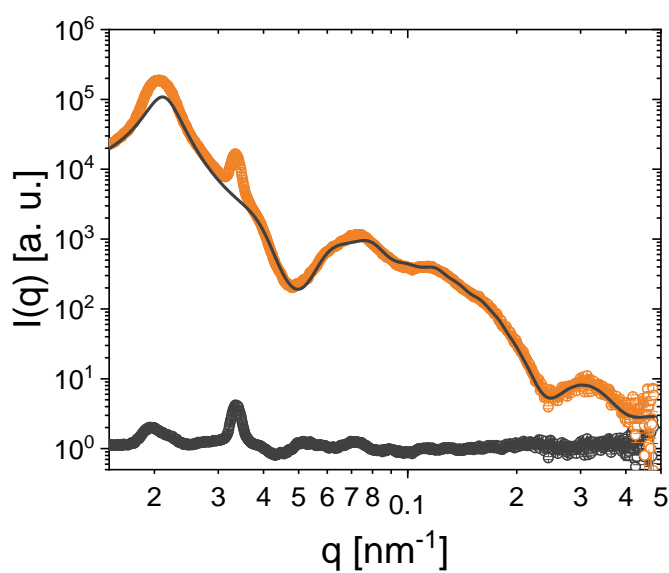

**Figure S22**. Scattering profile of the 5.4 wt% CS microgel dispersion (orange circles), the respective fit composed of the modeled form factor and the hard sphere structure factor (black line) and the residual structure factor associated to crystalline structures.

The extracted structure factor $S_{crystal}(q)$, related to the small crystallites only exhibits two Bragg peaks which in this case does not allow for a precise crystal structure analysis. Based on the CS microgel dispersions possessing higher mass contents we suppose closed packed structures like hcp or fcc.

**Lattice compression**

From the linear relationship between $q_{hkl}$, the position of the Bragg peak and the d-spacing of the crystal lattices we can extract the lattice constants a from the slope of the linear fit. From this we can calculate the lattice constant a. The slopes and respective lattice constants are listed in **Table S11**.

**Table S11**. Slopes and respective lattice constants extracted from linear fits of the Bragg peak position as function of the lattice spacing.

|  | 5.4 wt% | 7.3 wt% | 9.1 wt% | 10.9 wt% |
|---|---|---|---|---|
| Slope [nm$^{-1}$] | 0.0168 | 0.0190 | 0.0205 | 0.0219 |
| *a* [nm] | 374 | 331 | 306 | 287 |

We compare the lattice obtained from 2D simulations of the SAXS patterns and the linear relationship between the Bragg peak positions and $q_{hkl}$ in **Figure S23**. In addition, we also added the lattice constants calculated based on the Bragg peaks obtained from Vis-NIR absorbance spectroscopy. All extracted lattice constants are in good agreement with each other.

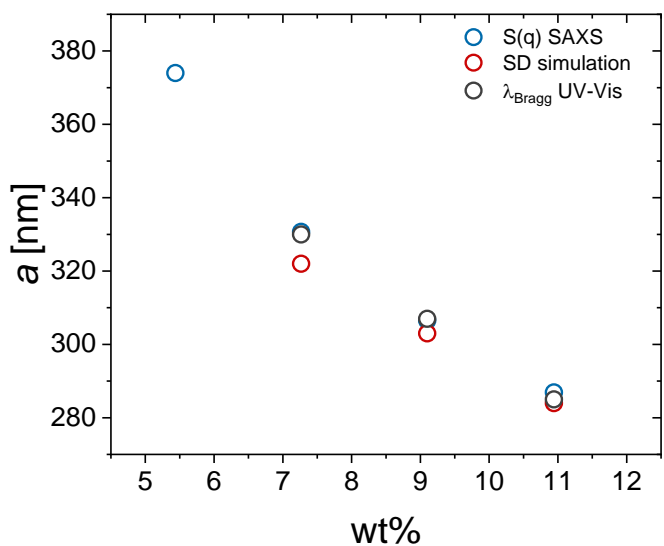

**Figure S23**. Lattice constants based on the hcp crystal structure extracted from $S(q)$, 2D simulation of the diffraction pattern and the Bragg peak from Vis-NIR absorbance measurements.

**Volume fraction of CS microgels in the colloidal crystals**

In addition to the volume fractions calculated based on $R_h$ as well as $R_{SAXS}$ normalized to $R_h$, we also calculated the volume fraction based solely on $R_{SAXS}$. Based on the scattering intensity of the $SiO_2$ cores we obtained the relation between particle number density and the mass content of the CS microgels. With the radius $R_{SAXS, dil.}$ of the CS microgels we calculated the generalized volume fraction illustrated as black line in **Figure S24**. In addition, we calculated the volume fraction $\phi_{crystal}$ of the CS microgels based on the lattice constant $a$ and $R_{SAXS, dil.}$ to compare $\xi_{generalized}$ and $\phi_{crystal}$ according to **equation S22**. We see a good agreement between $\xi_{generalized}$ and $\phi_{crystal}$ which shows the high reliability of the SAXS measurements conducted to extract particle number concentrations. When we include the isotropic osmotic deswelling of the CS microgels at dense packings we see a distinct deviation between the respective volume fractions and $\xi_{generalized}$ for mass contents above 7.3 wt%. The effective volume fractions of the CS microgels seems to stay constant around 0.43 where we identified the solid phases.

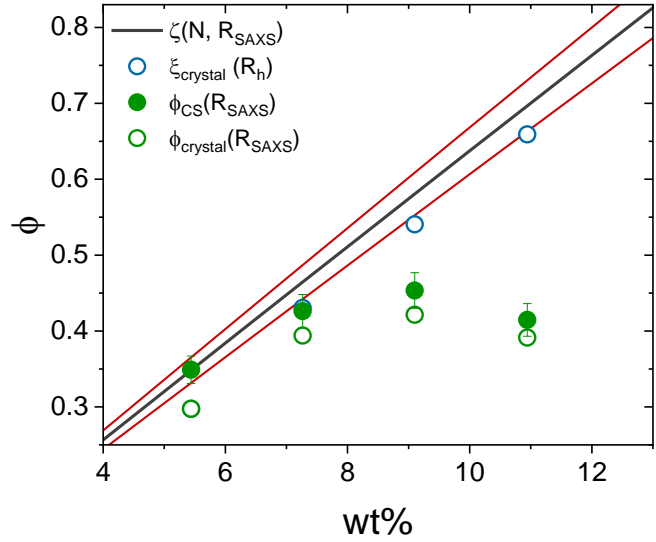

**Figure S24.** Relation between volume fraction and mass content of the CS microgels. The solid black line is related to the generalized volume fraction extracted from particle number concentration and $R_{SAXS, dil.}$ obtained from the P(q) of the dilute CS microgel dispersion. The red lines indicate the respective standard deviation. Green and blue circles are related to the volume fraction of the CS microgels based on the unit cell dimensions. For the blue circles the CS microgel volume is based on $R_{SAXS}$ from dilute dispersion, while for the green circles $R_{SAXS}$ is obtained from the respective modeled form factor in dense packing. The green dots indicate the volume fraction of the CS microgels based on the particle number concentration and the radius extracted from the respective modeled form factor in dense packing.

**Williamson-Hall analysis**

We performed a Williamson-Hall analysis to extract the radial and azimuthal sizes of the coherently scattering domains.[15] **Figure S25** shows the square of the FWHM of the Bragg peaks $w_{rad/azi}$ as function of the position of the Bragg peak in $q^2$. The FWHM and peak positions, regarding the radial domain sizes were extracted by the application of a Gaussian-fit-function on the Bragg peaks in the structure factor profiles. While for the determination of the azimuthal domain size we performed automated image analysis to apply gaussian fits on the azimuthal profiles of the respective Bragg peaks. A linear fit was applied on the obtained data according to:

$$w_{rad/azi}^2(q) = \left(\frac{2\pi}{L_{rad/azi}}\right)^2 + g_{rad/azi}\, q^2 \qquad (S29)$$

Here $L_{rad/azi}$ is related to the size of coherently scattering domains and $g_{rad/azi}$ is related to strains of the crystallites. $L_{rad/azi}$ can be extracted from the intercept of the linear fit and is presented in **Figure 25**. The extraction of g$_{rad/azi}$ leads to strains below 1% which is reflected by the small slope of the linear fits. From this we conclude that we have domain sizes of 3 to 4 µm with nearly no strains present for our colloidal crystals based on CS microgels.

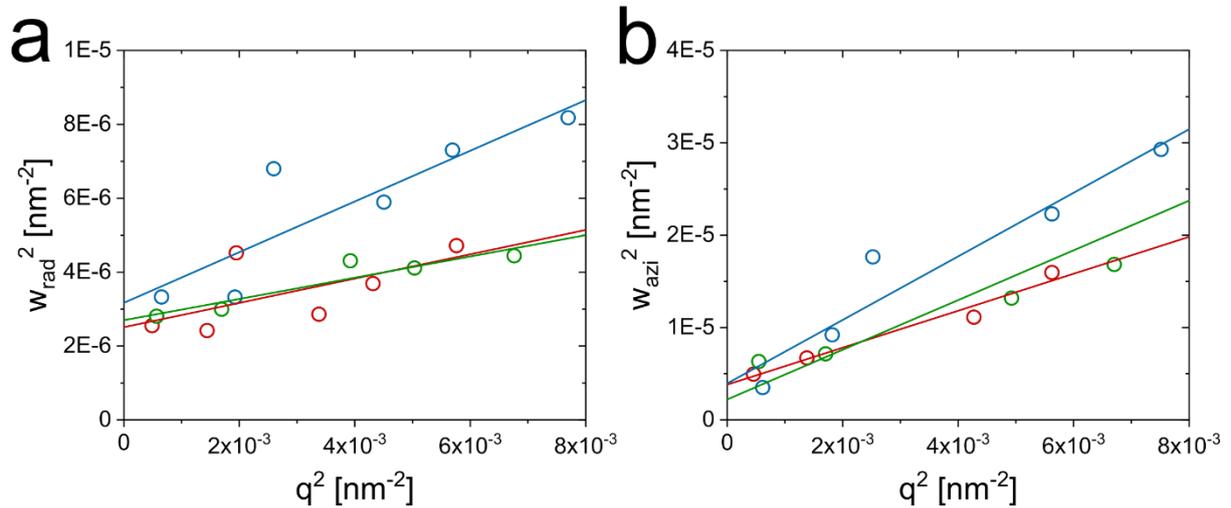

**Figure S25**. Williamson-Hall analysis applied on the Bragg peaks to extract the size of the coherently scattering domains of the colloidal crystals in (a) radial and (b) azimuthal sizes. Square of the FWHM from the Bragg peaks obtained from a gaussian fit as function of $q^2$. The solid lines correspond to linear fits to the data.

**Form factor modeling and radial density profiles of CS microgels at dense packing**

The distinct form factor oscillations in the scattering profiles of the CS microgel dispersions with high mass contents from 5.4 to 10.9 wt% enabled a detailed form factor analysis, based on the same core-exponential-shell model which was used to fit the scattering profile of the CS microgels in the dilute state. The parameters applied for the form factor modeling are listed in **Table S12**.

**Table S12**. Parameters used for the form factor modeling of the CS microgel dispersions in the concentrated regime.

| Parameters | 5.4 wt% | 7.3 wt% | 9.1 wt% | 10.9 wt% |
|---|---|---|---|---|
| scale | 0.28 | 0.34 | 0.36 | 0.39 |
| IB [a. u.] | 0.2 | 0.2 | 0.2 | 0.2 |
| $R_{core}$ [nm] | 18 | 18 | 18 | 18 |
| $\Delta t_{shell}$ [nm] | 120 | 116 | 109 | 98 |
| $SLD_{core}$ [$10^{-6}$ Å$^2$] | 17.75 | 17.75 | 17.75 | 17.75 |
| $SLD_{shell, in}$ [$10^{-6}$ Å$^2$] | 9.89 | 9.89 | 9.89 | 9.89 |
| $SLD_{shell, out}$ [$10^{-6}$ Å$^2$] | 9.43 | 9.43 | 9.43 | 9.43 |
| $SLD_{solvent}$ [$10^{-6}$ Å$^2$] | 9.43 | 9.43 | 9.43 | 9.43 |
| $\sigma_{core}$ | 0.1 | 0.1 | 0.1 | 0.1 |
| $\sigma_{shell}$ | 0.08 | 0.08 | 0.08 | 0.08 |
| A | 2.2 | 2.6 | 3.5 | 4.9 |

In addition, we were able to extract the radial density profiles of the CS microgels from the modeled form factor shown in **Figure S26a**. The density profile exhibits high ΔSLD values until reaching a radius of 18 nm, representing the SiO$_2$ core. From this point we see an exponential decrease in contrast ΔSLD which is related to the inhomogeneous structure of the microgel shell. With an increase in concentration the decline of the shell contrast gets more and more pronounced, also resulting in a decrease of the thickness of the shell. We integrated the regime of the density profile associated with the microgel shell for all profiles, presented in **Figure 26b**, in order to ensure constant mass and the validity of the ΔSLD profiles. Here we want to note, that the SLDs of SiO$_2$ and the PNIPAM microgel shell were obtained from the SLD calculator provided by NIST.[16]

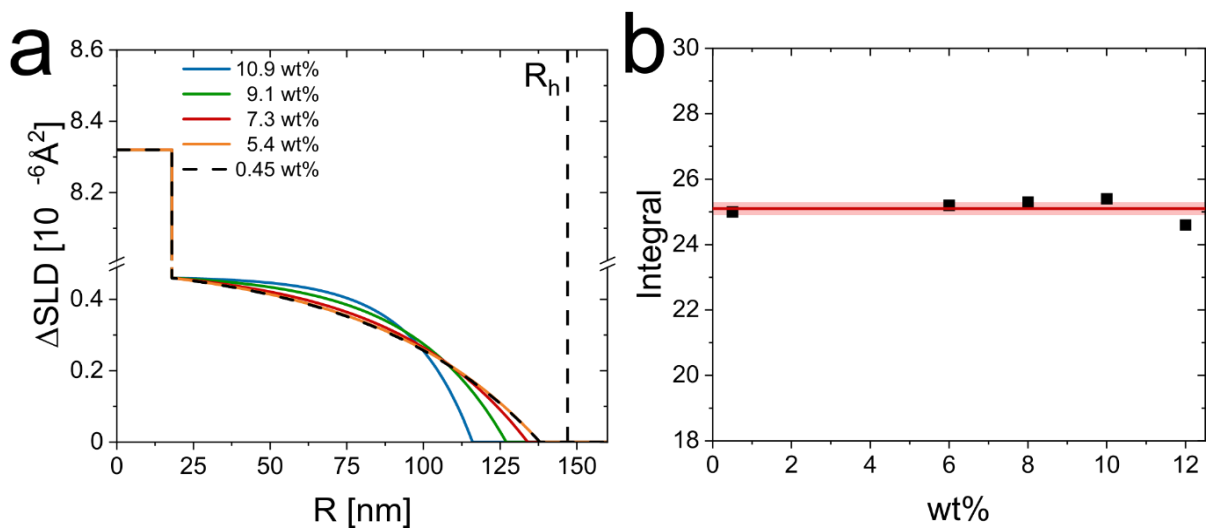

**Figure S26.** (a) Radial ΔSLD profiles extracted from the modeled and fitted form factors of the CS microgels in dense packings and dilute state. For the sake of clarity, we added a break on the y-axis, due to the strong differences in ΔSLD between the core and the shell. (b) The integrals of the microgel shell extracted from the radial ΔSLD profile in dependence of the respective mass content.